\documentclass[journal]{IEEEtran}
\usepackage{array}
\usepackage{float}
\usepackage{makecell}
\usepackage{placeins}
\usepackage{graphicx}
\usepackage{multirow}
\usepackage{amsmath}
\usepackage{mathtools}
\usepackage{epstopdf}
\usepackage{authblk}
\usepackage{subfigure}
\usepackage{cite}
\usepackage{cleveref}
\usepackage{latexsym}
\usepackage{amsthm}
\usepackage{balance}
\usepackage{relsize}
\usepackage{color}
\usepackage{amssymb}
\usepackage{eucal}
\usepackage{url}
\usepackage[utf8]{inputenc}
\usepackage{booktabs,amsfonts,dcolumn}
\usepackage{scalefnt}
\usepackage{geometry}
\usepackage[nodisplayskipstretch]{setspace}
\definecolor{Mypink}{RGB}{255,0,255}
\definecolor{Myorange}{RGB}{255,102,0}
\definecolor{Mygreen}{RGB}{0,153,0}
\definecolor{Myblue}{RGB}{0,0,255}
\newcolumntype{d}[1]{D..{#1}}

\DeclareGraphicsExtensions{.pdf,.jpeg,.png,.eps}
\begin{document}

\title{Optical Rate-Splitting Multiple Access for Visible Light Communications }

\renewcommand\Authfont{\fontsize{12}{14.4}\selectfont}
\renewcommand\Affilfont{\fontsize{9}{10.8}\itshape}
\author{Shimaa~Naser,
Lina~Bariah,~\IEEEmembership{Member,~IEEE},Wael~Jaafar,~\IEEEmembership{Member,~IEEE},
        Sami~Muhaidat,~\IEEEmembership{Senior~Member,~IEEE},
				Paschalis~C.~Sofotasios,~\IEEEmembership{Senior~Member,~IEEE},  
				Mahmoud~Al-Qutayri,~\IEEEmembership{Senior~Member,~IEEE}, and~Octavia~A.~Dobre, ~\IEEEmembership{Fellow,~IEEE}
				\thanks{S. Naser, S. Muhaidat, P. C. Sofotasios, and M. Al-Qutayri are with the Department of Electrical Engineering and Computer Science, Khalifa University, Abu Dhabi, UAE}
				\thanks{W. Jaafar is with the Department of Systems and Computer Engineering, Carleton University, ON, Canada}
				
			\thanks{O. A. Dobre is with the Faculty of Engineering and Applied Science University, Memorial University, St. John's, Canada}
				}

\maketitle

\begin{abstract}

The increased proliferation of connected  devices  and  emergence of internet-of-everything represent a major challenge for broadband wireless networks. This requires a paradigm shift towards the development of innovative technologies for the next generation  wireless systems. One of the key challenges towards realizing this vision, however, is the scarcity of spectrum, owing to the unprecedented broadband penetration rate in recent years.
A promising solution to the current spectrum crunch is the proposal of visible light communications (VLC), which explores the unregulated  visible light spectrum to  enable high-speed short range communications, in addition to providing efficient lighting.  
This  solution is envisioned to offer a considerably wider bandwidth, that can accommodate ubiquitous broadband connectivity to indoor users and further offload data traffic from overloaded cellular networks.
Although VLC is inherently secure and able to overcome the shortcomings of current RF wireless systems, it suffers from  several limitations, including the limited modulation bandwidth of  light-emitting diodes. In this respect, several interesting solutions have been proposed in  the recent literature to overcome this limitation. 
In particular, most common orthogonal and  non-orthogonal multiple access techniques initially proposed for RF systems, e.g., space-division multiple access and NOMA, have been considered in the context of VLC.  In spite of  their promising multiplexing gains, the  performance of these  techniques is somewhat limited. 
Consequently, in this article  a new and generalized multiple access technique, called rate-splitting multiple access (RSMA), is introduced and investigated for  the first time in  VLC networks. 
In this article,  we first provide an overview of the key multiple access technologies used in VLC systems. Then, we propose the first comprehensive approach to the integration of RSMA with VLC systems. In our  proposed  RSMA-VLC framework,  signal-to-interference and noise ratio expressions are derived and subsequently used to evaluate the weighted sum rate (WSR) performance of a two-user scenario. Our results illustrate the flexibility of RSMA in generalizing other multiple access techniques, namely NOMA and SDMA, as well as its superiority in terms of WSR  in the  context of VLC.\\
\end{abstract}

$\textup{\textbf{Index Terms}}$:  Multiple-input multiple-output, multiple access, NOMA, rate-splitting, SDMA, weighted sum rate, VLC.

\IEEEpeerreviewmaketitle
\section{Introduction} 
 
\begin{figure*}[t]
\centering
\includegraphics[width=1\linewidth]{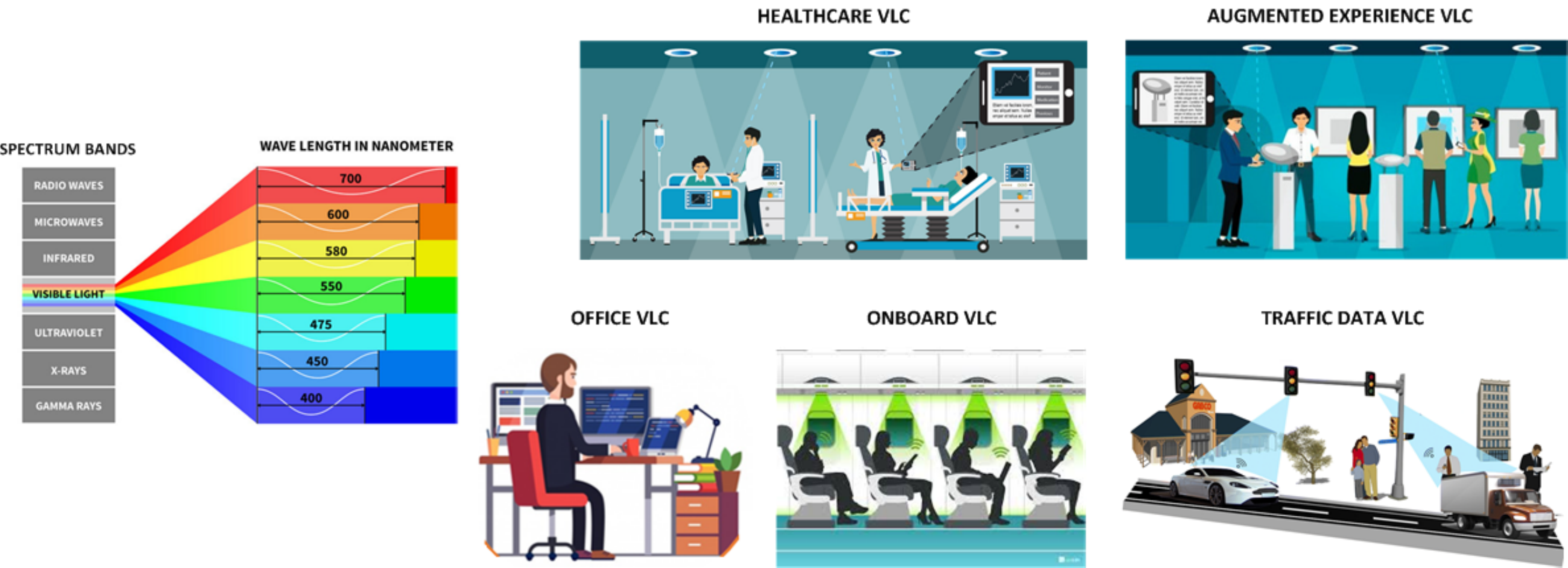}
\caption{VLC spectrum and use-cases.}
\label{Fig:VLC1}
\end{figure*}

 \IEEEPARstart{T}{he} exponential growth of connected devices and  emergence of the internet-of-everthing (IoE), enabling ubiquitous connectivity among billions of people and machines, have been the major driving forces towards the evolution of wireless technologies that can support a plethora of new services, including  enhanced mobile broadband and  ultra-reliable and low-latency communications (uRLLC). While the demand for new IoE services, e.g., eXtended reality
(XR) services, autonomous driving,  and tactile internet,  continues to grow, it is necessary for future wireless networks to deliver high reliability, low latency, and extremely high data rate. In this context, visible light communication (VLC) has emerged as a promising  wireless technology that is capable of providing high data rates and massive connectivity to users through fast fiber backhauling. To realize VLC, a very simple yet inexpensive  modification is required to existing lighting infrastructure \cite{Komine2004,Elgala2011,Ghassem2019}. The key attractive features of VLC include, but are not limited to security, high degree of spatial reuse, and immunity to electromagnetic interference (EMI) \cite{Grobe2013}.The advancement in solid-state has introduced light emitting diodes (LEDs) as  energy-efficient light sources, which are envisioned  to dominate the next generation wireless infrastructure. One of the interesting features of LEDs is their ability to rapidly switch between different light intensities in a way that is not perceptible to human eyes,  enabling them to be the main technology for VLC systems. The key principle of VLC is to use  emitted light from the LEDs to perform data transmission through intensity modulation and direct detection (IM/DD), without affecting the LEDs' main illumination function. 
The huge unregulated spectrum of visible light allows VLC to offload data traffic from RF/microwave systems while providing high data rates.  VLC uses the 400 THz to 789 THz visible light spectrum, which is characterized by  low penetration feature through objects,  secure VLC communications, and  high quality of service (QoS) in interference-free small cells designs \cite{Karuna2015,Mirami2015,IEEE2018}. 
Fig. \ref{Fig:VLC1} illustrates the used VLC spectrum band and some of its use-cases. 

\begin{table*}[t]
\scriptsize
\centering
\caption{List of acronyms and Abbreviations.}
\begin{tabular}[l] {|p{2.5cm}|p{6cm}|p{2.5cm}|p{5cm}|}
\hline
\textbf{Abbreviation }&\textbf{Definition}&\textbf{Abbreviation}&\textbf{Definition }  \\ \hline \hline 
ACO-OFDM& 	Asymmetrically Clipped Optical OFDM&MU 	&Multi-User  \\ \hline
ADO-OFDM&	Asymmetrically Clipped DC Biased Optical OFDM &MUD 	&Multi-User Detection\\ \hline
AWGN	&Additive White Gaussian Noise &
MUI	&Multi-User Interference\\ \hline
BD &	Block Diagonalization & NOMA &	Non-Orthogonal Multiple Access\\ \hline
BER &	Bit Error Rate &OCDMA 	& Optical Code Division Multiple Access \\ \hline

BC&	Broadcast Channel &OOC& optical orthogonal codes \\ \hline
CDMA &	Code Division Multiple Access &OFDM 	&Orthogonal Frequency Division Multiplexing \\ \hline
CD-NOMA&	Code Domain NOMA &OFDMA	&Orthogonal Frequency Division Multiple Access\\ \hline

CoMP 	&Coordinated multi-point &OMA & Orthogonal Multiple Access \\ \hline
CSI 	&Channel State Information &OOK	&On-Off Keying\\ \hline

CSIT 	&CSI at Transmitter &PD 	&Photo Detectors\\ \hline
CSK 	&Color Shift Keying &PD-NOMA& 	Power Domain NOMA\\ \hline
DC 	& Direct Current &PHY & Physical layer\\ \hline
DCO-OFDM& 	DC Biased Optical OFDM &QoS  &Quality-of-Service\\ \hline

DD 	&Direct Detection &RC	&Repeated Coding\\ \hline
FoV 	& Field of View &RGB	&Red, Green and Blue\\ \hline

ICI & Inter-Channel Interference & RSMA	&Rate Splitting Multiple Access \\ \hline
IM 	& Intensity Modulation  &SC &Super-position Coding\\ \hline
ISI 	&Inter-Symbol Interference &SDMA	&Space Division Multiple Access\\ \hline

LD 	& Laser Diode &SIC 	&Successive Interference Cancellation\\ \hline
LED 	&Light Emitting Diode &SM 	&Spatial Modulation\\ \hline
LTE & Long-Term Evolution &SMP	&Spatial Multiplexing \\  \hline

LoS 	&Line-of-Sight &SINR &Signal-to-Interference-Plus-Noise Ratio\\ \hline
MA	& Multiple Access &SNR & Signal-to-Noise Ratio \\ \hline
MAC & Media Access Control  &TDMA &	Time Division Multiple Access\\ \hline
MIMO &	Multiple-Input Multiple-Output&VLC 	&Visible Light Communication \\  \hline
MISO	&Multiple-Input Single-Output &WSMSE	&Weighted Sum Mean Squared Error\\  \hline
MMSE 	& Minimum Mean-Square Error&WSR	&Weighted Sum Rate\\ \hline
MSE 	&Mean-Square Error&ZF-DPC & Zero-Forcing Dirty-Paper Coding \\ \hline
 
%
\end{tabular}
\label{Table1}
\end{table*}

\subsection{Motivations and Contribution}
Despite its advantages, VLC suffers from several drawbacks that limit its performance. For example, the limited modulation bandwidth and peak optical power of LEDs are considered as the main obstacles towards realizing the full potential of VLC systems \cite{Jovicic2013}. Therefore, several studies  carried out to enhance the spectral efficiency of VLC systems. In particular, two research directions have been identified. With the former, researchers focused on the  design of dedicated VLC analog hardware and digital signal processing techniques. On the other hand, with the latter, researchers focused on enhancing the spectral efficiency through the development of different optical-based modulation and coding schemes, adaptive modulation, equalization, VLC cooperative communications, orthogonal and non-orthogonal multiple access (OMA/NOMA) schemes, and multiple-input-multiple output (MIMO) \cite{Pathak2015}.

In the context of VLC, several optical OMA schemes  have been proposed that include  time division multiple access (TDMA), orthogonal frequency division multiple access (OFDMA) \cite{Armstrong2009}, optical code division multiple access (OCDMA) schemes \cite{Salehi1989} and space division multiple access (SDMA). In TDMA, different users are allocated different time slots for communication, while in OFDMA, different users are assigned different orthogonal frequency sub-carriers. In  OCDMA,  users communicate at the same time/frequency, which can be achieved through  the use of different orthogonal optical codes. Finally, SDMA exploits the spatial separation between users to provide full time and frequency resources.


On the other hand, NOMA has been recently introduced as a spectrum-efficient multiple access scheme that allows different users to share the same time/frequency resources, leading to an enhanced spectral efficiency \cite{Ding2014}. NOMA is realized either by assigning different power levels to different users based on their channel gain  or  by different spreading sequences. The former is  called power-domain (PD)-NOMA, whereas the latter is referred as  code-domain (CD)-NOMA.  

Recent research results have shown that RSMA in MIMO-based RF systems outperforms other common multiple access schemes in terms of spectral efficiency \cite{Mao2018}. This is mainly due to the low or null correlation among multi-antenna radio channels, that allows efficient precoding at the transmitter. However, in VLC systems, MIMO channels are highly correlated, which  inevitably degrades the performance of linear precoding schemes. This has motivated the investigation of different receiver structures and precoding schemes aiming at mitigating the effect of channel correlation in MIMO VLC systems. It is worth noting that uplink VLC is impractical due to the low-power limitations of portable devices and the inconvenient light produced by end users. Consequently, RF or infrared communications can be good candidates  for uplink transmissions.

Motivated by the above discussion, we first review the state-of-the-art multiple access and MIMO transmission techniques, which have been proposed as efficient solutions to improve the spectral efficiency of VLC systems. Moreover, we suggest the newly emerged RF-based RSMA as a promising solution  and discuss its potential integration with VLC systems. To the best of our knowledge, this is the first work that investigates the efficient integration of RSMA with MIMO-VLC systems. Simulation results are presented to demonstrate the superiority of RSMA compared to other conventional multiple access schemes.

\textit{Notation:} Bold upper-case letters denote matrices and bold lower-case letters denote vectors. $(\cdot )^{T}$ denotes the transpose operation, $\mathbb{E}(\cdot)$ is the expectation operation, $|\cdot|$ is the absolute value operation, \textbf{\rm{I}} is the identity matrix, \textbf{0} is the zero matrix, tr$(\cdot)$ is the trace of a matrix, 
and $\mathcal{N}(0,\sigma^2)$ is a real-value Gaussian distribution with zero mean and variance $\sigma^{2}$. Let $\textbf{z}=[z_1,\ldots,z_Z]$ be a vector of length $Z$, then $L_1(\textbf{z})=\sum_{i=1}^Z |z_i|$ is the $L_1$ norm.


\section{VLC Background}\label{sec:background}
In this section, we provide an overview of VLC main components, adopted modulation schemes, and  VLC channel model. Then, we present the state-of-the-art literature on  MIMO in VLC, considering both single-user and multi-user cases. All used acronyms and abbreviations are summarized in Table \ref{Table1}.        

\subsection{VLC Components} 
\begin{figure}[t]
\centering
\includegraphics[width=1\linewidth]{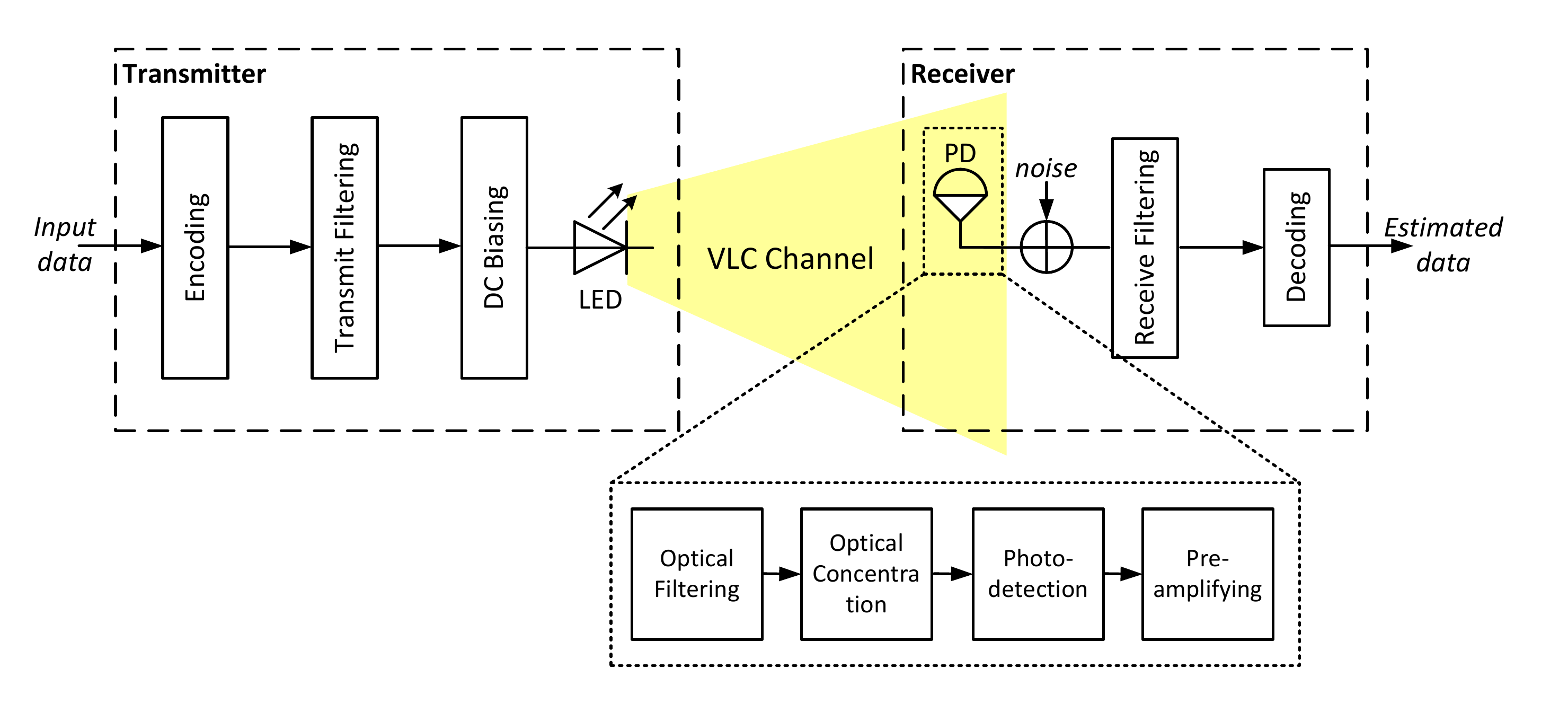}
\caption{VLC transceiver components.}
\label{Fig:VLCComp}
\end{figure}

The VLC components are illustrated in Fig. \ref{Fig:VLCComp}. An optical communication link is realized using an illuminating device at the transmitter to modulate  data using IM/DD. There are a variety of light sources that are available for optical communication,  but the most popular are LEDs and laser diodes (LDs). LEDs are the most popular illuminating devices with low fabrication costs. They are composed of solid-state semiconductor devices that produce spontaneous optical radiation when subjected to a voltage bias across the P-N junction \cite{Sze2007}. The direct current (DC) bias excites the electrons resulting in released energy in the form of photons. In most buildings, white LEDs are preferred since objects seen under white LEDs have similar colors to when seen under natural light. Two common designs are used for white LEDs. The first uses a blue LED with a yellow Phosphor layer, while the second combines red, green, and blue (RGB) LEDs. The first method is more popular thanks to its simplicity and low implementation costs. However, it suffers from  limited modulation bandwidth due to the intrinsic properties of the phosphor coating. On the other hand, RGB is more suitable for color shift keying (CSK) modulation, capable of achieving high data rate \cite{Monteiro2014}. LEDs offer many advantages over conventional illuminating devices such as florescent, incandescent and bulbs, which include  longer operational life time,  energy-efficiency, reaching up to 80\% compared to conventional devices,   very low radiation heat,  operate in extreme temperatures, easily directed and dimmed, and, lastly, very high switching rate \cite{Elgala2011}.

VLC receivers comprise   photo detectors (PD),  also called  non-imaging receivers or  imaging (camera) sensors, which are used to convert incident light power into electrical current proportional to light intensity.  A typical VLC receiver consists of an optical filter,  optical concentrator,  PD and  pre-amplifier. The optical filter eliminates  interference from  ambient light sources, while the optical concentrator enlarges the effective reception area of the PD without increasing its physical size. The optical concentrator is characterized by three parameters, i.e., field of view (FoV), refractive index, and radius. In order to increase the achievable diversity gain of an optical communication link, multiple receiving units can be deployed with different orientations,  optical filters, and concentrators. However, such  deployment comes at the expense of additional receiver size and complexity. To tackle this issue, an imaging sensor with a single wide FoV concentrator can be used to create multiple images of the received signals. Imaging sensors consist of an array of PDs that are integrated with the same circuit. It is worth noting that the required large number of PDs to capture high resolution photos renders them  energy inefficient.


\subsection{VLC Modulation Schemes} 
Conventional modulation schemes developed for RF systems cannot be readily applied to VLC channels. This is primarily due to the constraints of IM/DD which requires the
transmitted signal to be real and  positive, limiting the utilization of high order modulation. Motivated by this, the development of novel transmission strategies as well as efficient optical modulation techniques to enable high data-rate
VLC systems has attracted enormous attention in recent years. These modulation schemes range from simple on-off keying (OOK) modulation, where a “1” symbol is represented by a switched On LED state, and a “0” by a switched Off state \cite{Babar2017} to pulse position modulation (PPM) and pulse amplitude modulation (PAM). Both PPM and PAM provide higher data rates than OOK \cite{Lee2011,Wang2018}. Another method is to transmit data through  the variation of the colors emitted by LEDs, leading to the so-called color shift keying (CSK) modulation \cite{Monteiro2014}. CSK is similar to frequency-shift keying (FSK) in the RF domain. 
Nevertheless, the demand for better spectrally efficient schemes motivated the introduction of multiple-subcarrier modulations into VLC \cite{Carru1996}. For instance, orthogonal frequency division multiplexing (OFDM) is commonly used for its high spectral efficiency and ability to eliminate inter-symbol interference (ISI). However, its application to VLC is not straight-forward, as  the output of  LEDs have to be real-valued and positive \cite{Wu2015}.

Finally, efforts to standardize the PHY layer of VLC systems have led to the introduction of the IEEE 802.15.7 standard \cite{IEEE2018}.
Three physical (PHY) layers were released. Both PHY I and PHY II layers were allocated to a single light source and support mainly OOK and variable PPM (VPPM) modulation, where  PHY III layer considered  multiple light sources using the CSK modulation. 

\subsection{VLC Channel Model}

The PD's area of a  VLC system is much larger than the wavelength. Consequently, the multipath fading in an indoor VLC environment does not occur \cite{Dai2015,Marshoud2016}. Nevertheless, indoor optical links suffer from dispersion, modeled as a linear baseband impulse response. The latter can be assumed quasi-static since the mobility of users and connected objects is relatively low in indoor environments. 


\begin{figure}[t]
\centering
\includegraphics[width=200pt]{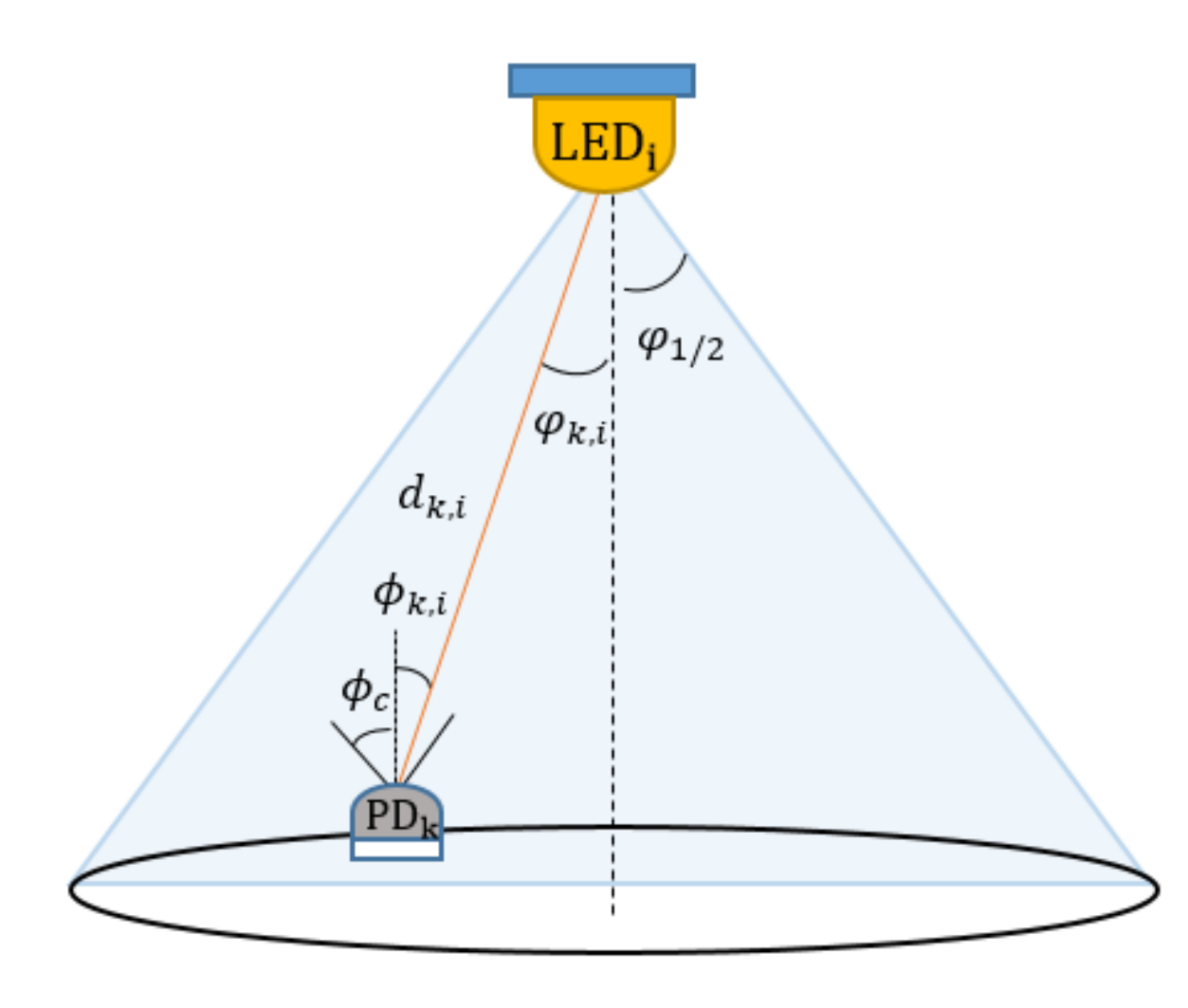}
\caption{VLC channel model (link between LED $i$ and PD $k$).}
\label{Fig:channel1}
\end{figure}

Typically, the channel of a VLC link can be modeled as follows. 
With the non-line-of-sight components neglected in front of stronger line-of-sight ones, the DC channel gain from the $i$th LED to the $k$th PD can be  expressed by {\cite{Komine2004}}
\begin{equation}
\label{eq:channel}
\small 
 h_{k,i}= \left\{\begin{matrix*}[l]
\frac{A_{k}}{d^2_{k,i}} R_{o}(\varphi_{k,i}) T_{s}(\phi_{k,i}) g(\phi_{k,i}) \cos(\phi_{k,i}),  & 0 \leqslant \phi_{k,i} \leqslant \phi_{c}\\
0, & \text{otherwise},
\end{matrix*}\right.
\end{equation}
where $A_{k}$ denotes the PD area, $d_{k,i}$ is the distance between the $i$th LED and $k$th PD, $\varphi_{k,i}$ is the angle of transmission from the $i$th LED to the $k$th PD, $\phi_{k,i}$ is the incident angle with respect to the receiver, and $\phi_{c}$ is the FoV of the PD. These angles are well-illustrated in Fig. \ref{Fig:channel1}. Moreover, $T_{s}(\phi_{k,i})$ is the gain of the optical filter, $g(\phi_{k,i} )$ is the gain of the optical concentrator, written as 
\begin{equation}
\label{eq:g}
g(\phi_{k,i})= \left\{\begin{matrix*}[l]
\frac{n^2}{\sin^2({\phi_{c}})},  & 0 \leqslant \phi_{k,i} \leqslant \phi_{c}\\
0, &  \phi_{k,i}>\phi_{c}
\end{matrix*}\right.
\end{equation}
where $n$ is the refractive index, and $R_{o}(\varphi_{k,i})$ is the Lambertian radiant intensity given by
\begin{equation}
\label{eq:R}
R_{o}(\varphi_{k,i})= \frac{m+1}{2\pi} \left(\cos(\varphi_{k,i})\right)^m,
\end{equation}
with $m$ the order of the Lambertian emission expressed by
\begin{equation}
\label{eq:m}
m = \frac{\ln{(2)}}{\ln\left({\cos(\varphi_{1/2})}\right)}, 
\end{equation}
where $\varphi_{1/2}$ is the LED semi-angle at half power. For a VLC link, the received noise can be modeled as a Gaussian random variable with zero mean and variance 
\begin{equation}
\label{eq:noise}
\sigma^{2} = \sigma^2_{\rm{sh}} + \sigma^2_{\rm{th}}
\end{equation}
where, $\sigma^2_{\rm{sh}}$ and $\sigma^2_{\rm{th}}$ are the variances of the shot and thermal noises, respectively. The shot noise is caused by the high rate of the physical photo-electronic conversion process. Its variance at the $k$th PD can be written as
\begin{equation}
\label{eq:noise2}
\sigma^{2}_{k,\rm{sh}} = 2qB\left(\zeta_k h_{k,i}x_{i}+I_{\rm{bg}}I_{2}\right), 
\end{equation}        
where $q$ is the electronic charge, $\zeta_k$ is the detector responsivity, $x_{i}$ is the transmitted signal by the $i^{th}$ LED, $B$ is the corresponding bandwidth, $I_{\rm{bg}}$ is the background current, and $I_{2}$ is the noise bandwidth factor. Whereas, the thermal noise is resulting from the transimpedance receiver circuitry. Its variance at the $k^{th}$ PD is given by
\begin{equation}
\label{eq:noise3}
\sigma^{2}_{k,\rm{th}} = \frac{8\pi K T_{k}}{G}\eta A_{k} I_{2} B^{2} + \frac {16 \pi^{2} K T_{k} \gamma}{g_{m}} \eta^{2} A^2_{k} I_{3} B^{3}
\end{equation}
where $K$ is Boltzmann's constant, $T_{k}$ is the absolute temperature, $G$ is the open-loop voltage gain, $A_k$ is the PD area, $\eta$ is the PD's fixed capacitance per unit area, $\gamma$ is the field-effect transistor (FET) channel noise factor, $g$ is the FET transconductance, and $I_{3}$  = 0.0868 \cite{Komine2004}.
Modern infrastructures are commonly equipped with LED fixtures or arrays. A single fixture is composed of $Q$ LEDs, and may be viewed as a single VLC source\footnote{In the remaining of this paper, we interchangeably designate by LED a fixture of LEDs.}, where the DC channel gain may be expressed as
\begin{equation}
\label{eq:fixture}
\small 
h_{k,j}=\left\{
\begin{array}{ll}
A_k \sum \limits_{i=1}^Q d_{k,j,i}^{-2} R_o(\varphi_{k,j,i}) T_s(\phi_{k,j,i}) g(\phi_{k,j,i}) \cos(\phi_{k,j,i}),  \\
\mbox{if} \; 0 \leq \phi_{k,j,i} \leq \phi_c \\
0, \; \mbox{otherwise}\end{array},
\right.
\end{equation}
where $d_{k,j,i}$ is the distance between the $i$th LED in the $j$th fixture and $k$th PD, $\varphi_{k,j,i}$ is the angle of transmission from the $i$th LED in the $j$th fixture to the $k$th PD, and $\phi_{k,j,i}$ is the incident angle with respect to the receiver. Since the separation between LEDs in the same fixture is negligible compared to the distance between the fixture and $k$th PD, then distances and angles implicating index $i$ can be assumed approximately the same for all LEDs. Hence, the channel gain from the $j$th fixture to the $k$th PD can be given by
\begin{equation}
\label{eq:fixture_app}
\small 
h_{k,j}\approx \left\{
\begin{array}{ll}
Q \; h_{k,i},  & 0 \leq \phi_{k,j,i} \leq \phi_c, \; \forall i  \\
0,  & \mbox{Otherwise}\end{array}.
\right.
\end{equation}

\subsection{Single-user MIMO VLC}
Recalling that one of the main drawbacks of using LEDs for data transmission is their limited modulation bandwidth, MIMO techniques have been proposed for VLC in order to improve the spectral efficiency. Moreover, MIMO enables accurate physical alignment between transmitting LEDs and receiving PDs \cite{Zeng2009}. MIMO channels in VLC are known to be highly correlated, since there is no phase or frequency information in the IM/DD modulation \cite{Elgala2011}. In order to achieve high rank MIMO channels in VLC, two different configurations for the PDs’ orientation were designed in \cite{Ghassem2019}, namely, pyramid receiver and hemispheric receiver. In \cite{Nuwan2015}, the authors proposed a novel angle diversity receiver for MIMO VLC. The reported results showed that this receiver design achieves better performances, in terms of channel capacity and bit error rate (BER), than only spatially separated receivers.  

Different transmission techniques have been proposed for MIMO VLC. For instance, repetition coding (RC) is the simplest, where the same signal is transmitted by different LEDs. RC offers an improved BER performance. However, due to the high channels correlation in VLC systems, it requires large signal constellation sizes to achieve high data rates \cite{Fath2013}. In contrast, spatial multiplexing (SMP) has shown to enhance the spectral efficiency, by allowing different signal transmissions from multiple neighboring LEDs. Due to the high correlation between VLC channels, SMP experiences inter-channel interference (ICI). Consequently, different MIMO precoding techniques have been proposed to overcome this issue. Authors in \cite{Fath2013} investigated the performances of a MIMO VLC system where spatial modulation (SM) is combined with SMP. By definition, SM adds an extra dimension, i.e., the spatial dimension, to the signal constellation diagram. The spatial dimension associates a unique transmitter LED index with each binary data sequence.
By doing so, SM is able to avoid ICI, besides improving the spectral efficiency performance \cite{Mesleh2011,Cai2016}.
Another approach to alleviate ICI in SMP is the design of collaborative constellation (CC) \cite{Zhu2015}. In CC, symbols of different transmitting LEDs are jointly generated in order to minimize the average transmit optical power for a given minimum Euclidean distance. Through simulations, it has been shown that CC outperforms SMP and SM in terms of BER.

\subsection {Multi-user MIMO VLC}
MIMO transmissions have been also investigated for the multi-user scenarios, where a transmitter  sends different signals to spatially spread receivers at the same time. This can be realized by assuming a collocated set of LEDs that communicates simultaneously with multiple users in an indoor environment. 
Initially, multi-user MIMO (MU-MIMO) was proposed for use in the long-term-evolution (LTE) and LTE-advanced RF standard releases 8 and 10 \cite{Liu2012,Lim2013}. Its interesting performance in terms of increased capacity and coverage motivated its application into VLC. Researchers recently proposed several adaptations of MU-MIMO into VLC systems, as summarized in \cite{Ahmadi2018}.

In VLC systems, channel access can be realized either through multiple access channel (MAC) or broadcast channel (BC). Most research focused on the analysis of the downlink BC of MU-MIMO, with an emphasis on the data rate performance. These systems experience interference when orthogonal frequency/time resources are limited.  

Similar to ICI, multi-user interference (MUI) is a common issue in MU-MIMO systems. MUI can be eliminated at the receiver using an efficient multi-user detection (MUD) technique \cite{Spencer2004,Spencer2004_2}. However, implementing MUD in VLC systems suffers from high complexity and energy inefficiency. Consequently, most research in VLC focused on designing  data precoding techniques at the transmitter. For instance, block diagonalization (BD), which is a generalized form of channel inversion precoding, was introduced in \cite{Chen2013}. Although BD is a simple linear precoding technique, its application is limited to the scenario where the number of transmitting LEDs is larger than the total number of served users.

The authors in \cite{Hong2013} used BD precoding in downlink MU-MIMO VLC and showed that BD is limited by the correlation of the users’ channels. Hence, a scheme that reduces this correlation was proposed, based on the adjustment of PDs' FoVs. Authors in \cite{Yu2013} developed a linear zero-forcing (ZF) and a ZF dirty paper coding (ZF-DPC) schemes in order to eliminate MUI and maximize max-min fairness or throughput. However, authors of \cite{Ma2013} relaxed the ZF condition by applying the minimum mean squared error (MMSE) as a performance metric for the precoder design, for both perfect and imperfect CSI scenarios. In \cite{Li2015}, an optimal MSE precoder was designed in order to minimize the BER, under per-LED power constraints. The transceiver design was later simplified by adopting a ZF precoder. Results show that the simplified scheme outperforms the conventional ZF precoder in terms of BER, while MSE achieves the best performance. Similar designs were proposed in \cite{Cheng2016,Pham2017}, where an optimal ZF precoder was obtained using an iterative concave-convex procedure, aiming at maximizing the achievable per-user data rate. Then, the authors simplified the precoder design using the high signal-to-noise ratio (SNR) approximation.  In \cite{Shen2016_2}, the authors proposed a different beamforming technique aiming at maximizing the sum rate of a virtual MIMO VLC system. Beamforming was designed using the sequential parametric convex approximation (SPCA) method, and it has been shown through simulations that it outperforms conventional ZF-based beamforming, specially for highly correlated VLC channels and low optical transmit powers. Authors of \cite{Wang2015} proposed precoding for an OFDM-based MU-MIMO VLC system, where precoding is applied at each sub-carrier, using ZF and MMSE techniques.
This led to the enhancement of the sum rate performance at high SNR and for uncorrelated channels. The work in \cite{Ma2015,Ma2018} focused on precoding designs for coordinated multi-point (CoMP) MU-MIMO VLC systems. Through numerical analysis, authors illustrated the realized improvements in terms of signal-to-interference-plus-noise ratio (SINR) and Weighted sum mean square error, respectively. 

Since the transmitted signals in VLC systems are restricted to be positive and real, capacity analysis under Gaussian channel is different from that of RF systems. Several comprehensive analytical models were proposed in the literature to quantify the VLC channel capacity under transmit power constraints. For instance, the authors of \cite{Moser2017,Moser2017_2} derived high SNR asymptotic capacity expressions for MIMO VLC channels. They extended their results in \cite{Moser2018} to upper and lower bounds at both low and high SNRs. In \cite{Chaaban2018}, the authors derived low-SNR asymptotic capacity of a MIMO VLC channel, under both average and peak intensity constraints. Finally, \cite{Zhou2019} proposed new outer and inner capacity region bounds under per-user average or peak power.

\section {Multiple Access in VLC} \label{sec:MA}
\begin{figure*}[t]
\centering
\includegraphics[width=0.85\linewidth]{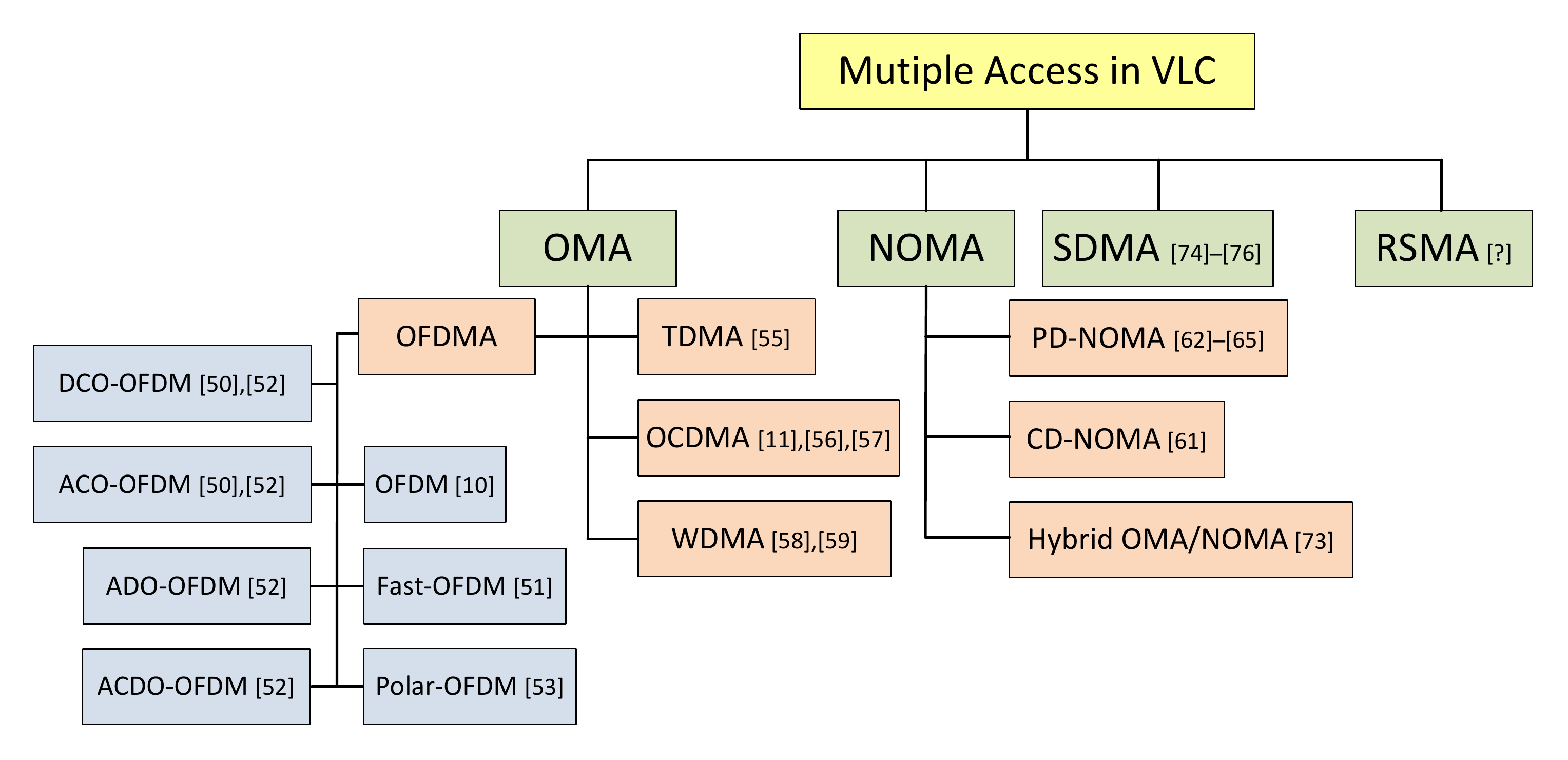}
\caption{Multiple Access in VLC: Related Work}
\label{Fig:Related}
\end{figure*}

Inspired by promising multiplexing gains achieved by conventional multiple access techniques, which were developed for RF systems, optical multiple access have received great attention recently. In this section, we summarize these schemes and the underlying approaches to adapt them into  VLC. A related work diagram is also presented in Fig. \ref{Fig:Related}.

\begin{figure*}[t]
\centering
\includegraphics[width=1\linewidth]{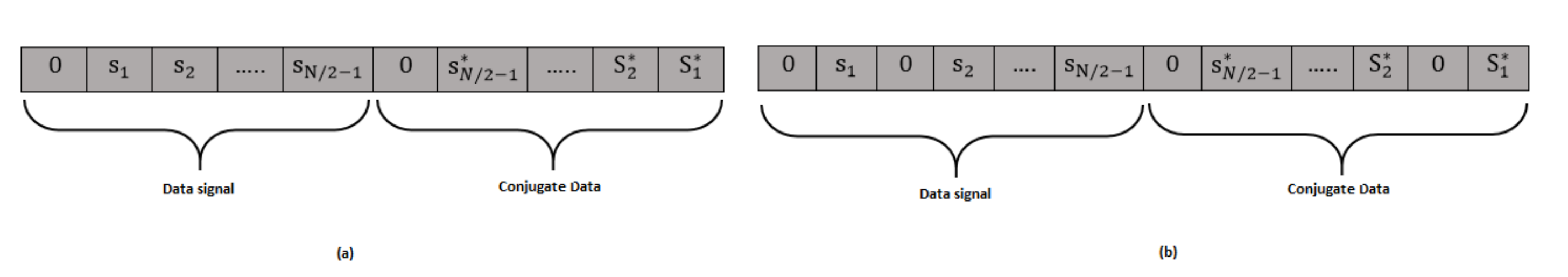}
\caption{Hermitian Symmetry in OFDM \cite{Ahmadi2018}. (a) DCO-OFDM, (b) ACO-OFDM.}
\label{Fig:Hermitian}
\end{figure*}
\subsection {Orthogonal Multiple Access}
In OMA systems, users are allocated orthogonal  frequency/time resources. Due to the limited modulation bandwidth of LEDs, high data rate transmissions result in ISI, which can significantly degrades the performance.  A key solution to eliminate ISI is to use OFDM.  However, due to the constraint of real-valued and positive modulated symbols, OFDM cannot be directly applied to VLC systems. To obtain real-valued signals, parallelized data streams are processed through  Hermitian symmetry before the inverse fast fourier transform (IFFT) operation, leading to 50\% loss in spectral efficiency. Furthermore, to ensure positive optial signals and realize IM/DD, a DC bias is added to the output of the IFFT operation\cite{Wu2015}. 

Several optical OFDM techniques were proposed, such as DC biased optical OFDM (DCO-OFDM), asymmetrically clipped OFDM (ACO-OFDM), asymmetrically clipped DC-biased optical OFDM (ACDO-OFDM), fast-OFDM and, polar-OFDM (P-OFDM) \cite{Armstrong2009,Mesleh2011_2,Giaco2012,Dissan2013,Elgala2015}. Fig. \ref{Fig:Hermitian} illustrates the Hermitian symmetry operation of DCO-OFDM and ACO-OFDM \cite{Ahmadi2018}, where $s_i$ is the $i$th signal and $s_i^*$ is the conjugate symmetric signal. As a main drawback, OFDM suffers from  high peak-to-average power ratio, which is   difficult to overcome in VLC systems due to the non-linearity of LEDs. Optical OFDM techniques can be extended to  multiple user scenaros, resulting in the so called  OFDMA.  OFDMA is realized by assigning different orthogonal frequency sub-carriers to different users, as illustrated in (Fig. 4, \cite{Kizi2015}). 

Time-division multiple access  is another common multiple access schemes, which allows different users to share the same frequency resources at different non-overlapping  time instances. However, its application in multi-user VLC systems is limited due to power and synchronization complexity constraints \cite{Jenq2011}.  

Optical CDMA has been proposed as an extended version of conventional CDMA to  optical systems \cite{Salehi1989,Salehi1989_2}. In OCDMA, multiple transmissions occur at the same time over the same frequency band, where signals are separated using optical orthogonal codes (OOC). Although OCDMA achieves better spectral efficiency than OFDMA, it requires long OOC sequences that reduces the achievable data rate. Moreover, it suffers from high complexity and non-ideal cross-correlation \cite{Rashidi2013}.

Similar to OFDMA, wavelength division multiple access (WDMA) has been proposed to achieve orthogonal MA \cite{Wang2015_2,Cossu2015}. In this scheme, multi-color LEDs are used to allow simultaneous transmissions at different wavelengths. In spite of its advantages, the implementation of dense-WDMA using the current LED technology is very complex and difficult to achieve. 

To summarize, OMA schemes efficiently mitigate  interference among users' signals by allocating orthogonal resources.  However,  the number of served users is limited and cannot exceed the number of available orthogonal resources. This concern is also true for VLC systems. 
Motivated by this, researchers have recently focused  on the design of novel non-orthogonal MA techniques, as explained in the next subsection.

\subsection {Non-Orthogonal Multiple Access}

\begin{figure}[t]
\centering
\includegraphics[width=220pt]{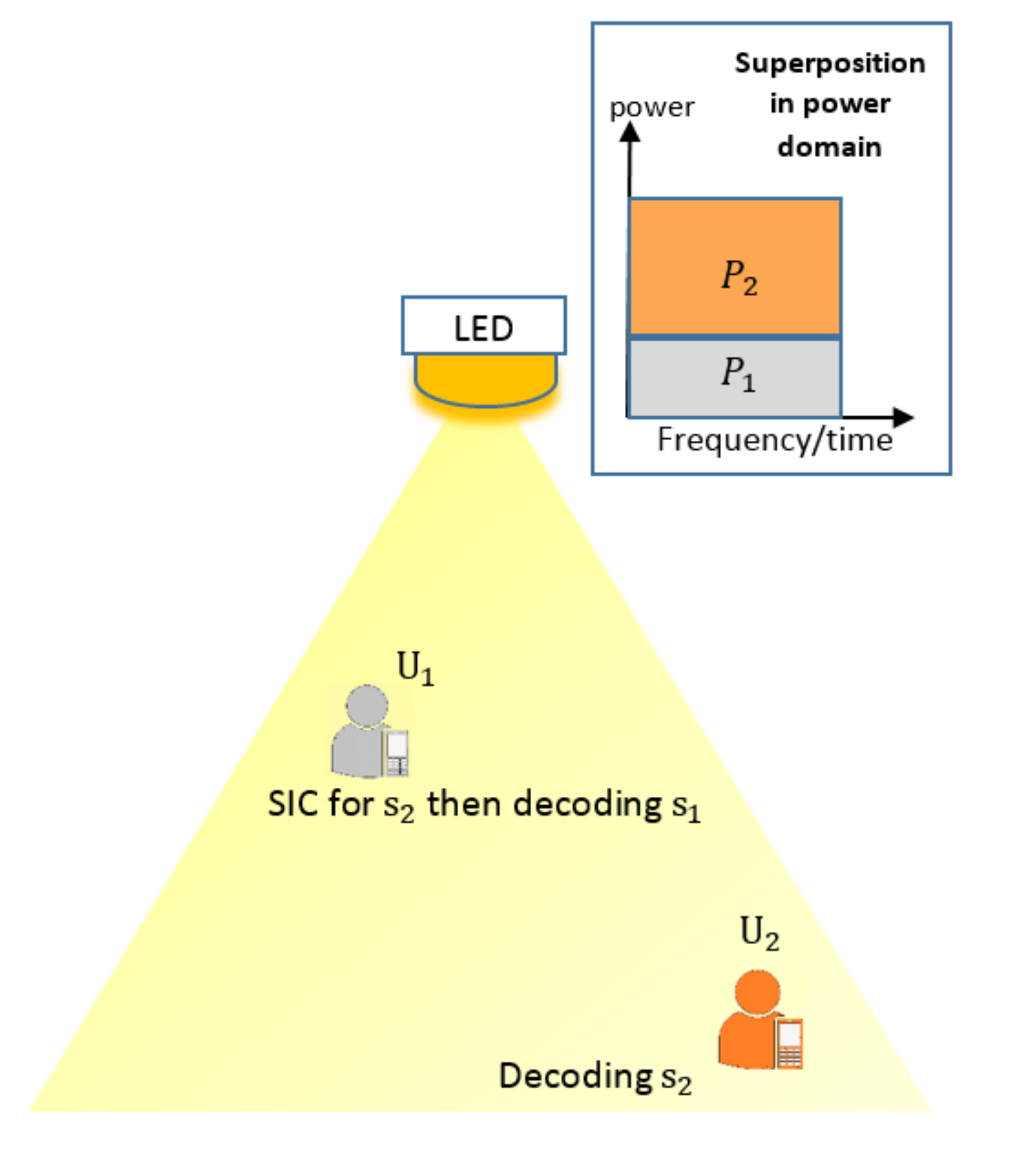}
\caption{Scenario of two users PD-NOMA in VLC.}
\label{Fig:PDNOMA}
\end{figure}

NOMA has emerged as a promising candidate to enhance spectral efficiency in 5G networks \cite{Ding2014}. The key principle of NOMA is to allow different users to share the same frequency resources simultaneously at the expense of MUI. To perform MUD, different users are assigned distinct power levels, which is referred to as PD-NOMA, or different spreading sequence, known as CD-NOMA \cite{Ding2014,Ali2017,Wei2018}. In PD-NOMA, successive interference cancellation (SIC) is utilized to eliminate users' signals with higher power levels, before detecting the intended user's signal. A basic system model for the two-user PD-NOMA is illustrated in Fig. \ref{Fig:PDNOMA}. First, users are ordered based on their channel gains. Then, the user with the weakest channel gain is allocated the highest amount of power, e.g., signal $s_2$ for user $U_2$. On the other hand, the user with the best channel gain is allocated the lowest power level to transmit its signal, e.g., signal $s_1$ for user $U_1$. Each user performs SIC in order to isolate and detect its own signal, except for the one with the weakest channel gain, where signal detection is performed considering interference from other users as noise.

Studies have shown that PD-NOMA can be efficiently applied  to VLC for the following reasons. First, PD-NOMA depends highly on  CSI, which can be readily available in VLC scenarios. Second, PD-NOMA performs best at high SNR values, which is mostly the case for VLC channels. Third, NOMA performs efficiently when users’ channels conditions are independent. This can be  achieved by changing the angle of the transmitting LED and the PD's FoV \cite{Marshoud2017}. Furthermore, reported results  showed that NOMA outperforms OFDMA and TDMA in terms of system capacity and number of simultaneously served users \cite{Kizi2015,Yin2016,Yang2017,Shen2017}. Although NOMA is suitable for cases, where the number of users is higher than the number of available orthogonal resources, its complexity grows rapidly and proportionally to the number of users, since the $k$th user needs to decode the messages of the $k-1$ users before detecting its own signal. To address this issue, a simple approach is to group users into small clusters, such that users of the same cluster communicate using NOMA, while the different clusters are scheduled using an OMA technique. It is worth mentioning that NOMA achieves interesting performances as long as users experience channels with different gains. 

MIMO techniques can efficiently enhance NOMA systems' performance by exploiting the availability of multiple transmitters. One of the two following approaches can be adopted. In the first approach, superposition coding-SIC (SC-SIC) is applied such that all users are sorted based on their effective precoded channels \cite{Wyner1974,Nguyen2017}. However, in the second, a number of users are paired into one cluster. In this approach, clusters must be separated using a SDMA technique in order to reduce inter-group interference \cite{Zeng2017_3}. According to results in \cite{Zeng2017_3,Zeng2017_2}, MIMO-NOMA outperforms MIMO-OMA in terms of sum rate and user fairness. Despite the aforementioned advantages of MIMO-NOMA systems, they come at the expense of a complex transmitter design, where joint optimization of precoding/decoding orders is required for different users.  

As explained earlier,  MIMO design can be realized by assuming multiple transmitting LEDs and multiple PDs at the receiver. Such system cannot employ the same power allocation method designed for  single transmitting LED NOMA VLC systems, such as gain ratio power allocation (GRPA) \cite{Marshoud2016}. Accordingly, several power allocation strategies have been proposed in the literature for MIMO NOMA, e.g., hybrid precoding and post-detection \cite{Ding2016}, and signal alignment \cite{Ding2016_2}. However, their counterpart in the NOMA-based MIMO VLC is almost non-existent. To our knowledge, only the authors of \cite{Chen2018} investigated NOMA-based MIMO VLC systems and proposed a power allocation strategy, called Normalized Gain Difference Power Allocation (NGDPA). The reported results for NGDPA illustrate a sum rate improvement of 29.1\% compared to GRPA.   Finally, authors in \cite{Abumarshoud2019} proposed a hybrid OMA/NOMA scheme for VLC, where the Downlink transmitter is smart enough to select dynamically the adequate MA technique according to the environment conditions. 



\subsection {Space Division Multiple Access} 
In recent buildings design, it is common to have multiple illuminating LEDs in indoor spaces. This configuration motivated the deployment of SDMA. In this scheme, transmit angle diversity is used to create narrow-band beams towards sparsely located users, while achieving the same coverage of a single wide-beam transmitter \cite{Carrut2000,Kim2014,Chen2015}. The advantage is that more power is directed into each user; hence improving the communication's reliability. While SDMA in RF systems requires a complex beamforming technique to create the narrow-band beams, it is much simpler in VLC, where narrow-band beams can be achieved by reducing the FoVs of LEDs. In order to avoid interference among  users, spatial separation needs to be implemented by adequately allocating transmit power among beams directed to users. Consequently, each receiver attempts to detect its signal while treating any interference as noise.


Although SDMA renders the transmitter and receiver design simpler compared to NOMA, it becomes inefficient as soon as the number of users becomes larger than the number of transmit LEDs, i.e., an overloaded scenario. It should be noted that the number of LEDs has to be more than or  equal to the the number of users in order to guarantee interference reduction. Moreover, due to real-valued signals in VLC, it is very difficult to pair orthogonal users together, as in the case of RF. Hence, the performance of SDMA   degrades in VLC scenarios. 


\begin{figure*}[t]
\centering
\includegraphics[width=1\linewidth]{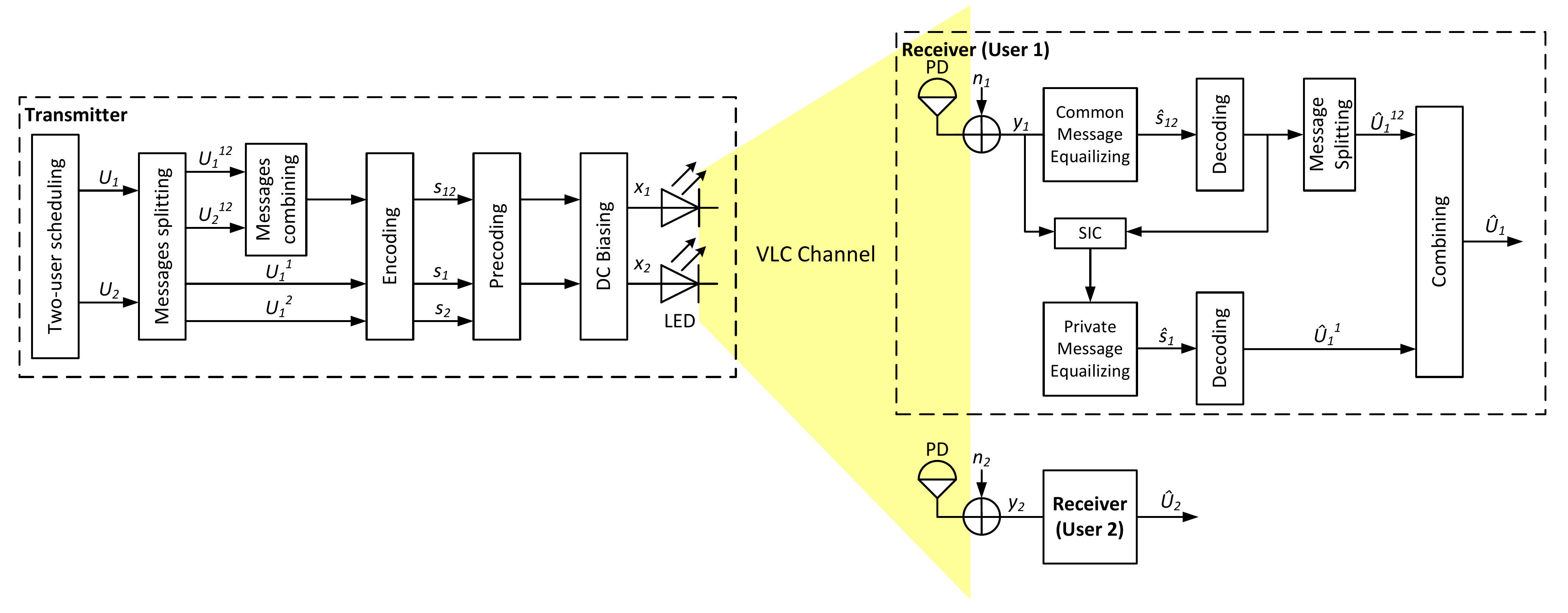}
\caption{RSMA-based two-user MISO VLC system.}
\label{Fig:2MISO}
\end{figure*}

\section {Rate-Splitting Multiple Access}

In the previous section, we summarized the main OMA and NOMA techniques and their integration with VLC systems. While NOMA enables simultaneous transmissions to a large number of users, i.e., overloaded scenario and SDMA achieves spatial separation between users in underloaded scenarios, a generalized configuration can be exploited in order to optimize the utilization of resources for any scenario. This motivated the proposal of RSMA as a generalization scheme, where NOMA and SDMA can be considered as special cases. 

\subsection {Background}

In \cite{Mao2018}, RSMA was proposed as a powerful and generalized MA technique for RF systems. It is envisioned that RSMA will achieve tactile improvements in the field of MA techniques, by allowing the network to efficiently serve multiple users with different capabilities in overloaded and underloaded scenarios. According to the key principle of RSMA, which relies on the implementation of  linear precoding at the transmitter and SIC at the receiver, it is expected to fill the gap between NOMA and SDMA techniques. 
In RSMA, users' messages are split into common and private parts at the transmitter. Then, a combiner is used to multiplex the common parts of all users and encode them into a single common stream. Meanwhile, the private parts are encoded separately into multiple private streams. Afterwards, a linear precoder is used to mitigate MUI. Finally, all precoded streams are superimposed on the same signal and sent over a VLC BC channel. At each user, the common stream is decoded and the user's intended data is extracted. Then, interference introduced by the common stream is eliminated using  SIC, as in NOMA. Subsequently, the private part of each user message can be decoded, while treating the private parts of other users messages as noise, as in SDMA. This mechanism is illustrated in (Fig. 1, \cite{Mao2018}). 

RSMA depends mainly on the splitting design of  messages and  power allocation strategies between  common and private parts of users' messages. Extensive research efforts were devoted to the investigation of these issues in order to improve the  efficiency of RSMA \cite{Mao2018,Hao2015,Dai2016,Dai2017,Joudeh2017,Papaz2017}. In \cite{Mao2018}, the authors provided a solid analytical framework to study the performance of RSMA in multi-user multiple-input-single-output (MU-MISO) BC channels. The reported results proved that RSMA outperforms NOMA and SDMA in terms of sum-rate for different users' setups. In \cite{Hao2015}, the authors proposed a practical scheme for private symbols encoding in RSMA using the conventional ZF beamforming. Then, they 
studied the sum-rate performance for a two-user BC channel with limited CSI feedback  (CSIT). The authors in \cite{Dai2016} investigated RSMA with massive MIMO and  imperfect CSIT. 
They proposed a hierarchical-rate-splitting (HRS) framework where two different types of common messages are defined, that can be decoded by either all users or by a subset of them. Then, the associated sum rate performance was investigated in order to adjust the precoders of common messages. Numerical results illustrated the superiority of HRS compared to conventional techniques such as TDMA and BC with user scheduling. This work was extended in \cite{Dai2017} to a multi-user mmWave case, where CSIT is either statistical or quantized. Similar to \cite{Dai2016}, the authors in \cite{Joudeh2017} proposed a hybrid messages RSMA precoding in order to achieve max-min fairness amongst multiple co-channel multicast groups. The superiority of their approach is proved through degree-of-freedom (DoF) analysis and simulations. Finally, \cite{Papaz2017} evaluated the robustness of RSMA, given the presence of hardware impairments such as phase, distortion and thermal noises, and the availability of perfect/imperfect CSIT. 

Despite the extensive research efforts on RSMA in the RF domain, to the best of our knowledge, its use in VLC systems has not been proposed nor analyzed. Therefore, we provide in this paper the first comprehensive approach for analyzing RSMA in  VLC systems. In particular, we develop a framework that analyses the performance of RSMA and compares its performance with existing VLC techniques. Furthermore, we give insights into the challenges and future research directions for RSMA-based VLC systems.

\subsection {System Model} \label{sec:models}
RSMA was initially proposed for RF communication systems in \cite{Elgamal2011}, as a method of achieving a new rate region for  two-user BC channels. The concept of RSMA was extended to a multi-antenna BC channel   in order to bridge the gap between two extreme multiple access schemes, namely NOMA and SDMA \cite{Mao2018}. The authors of \cite{Mao2018} have shown that RSMA works best in the  multiple-input case. In the VLC context, this can be seen as using several transmitting LEDs to create a BC channel towards users.   


Hence, we analyze in this paper the performance of RSMA in downlink BC VLC networks. For the sake of simplicity, we assume two transmitting LEDs, that  send messages to two single-PD users, as depicted in Fig. \ref{Fig:2MISO}. Messages $U_1$ and $U_{2}$ are intended to users 1 and 2, respectively. $U_1$ is divided into two parts, private part $U^1_{1}$ and common part $U^{12}_{1}$. Similarly, $U_2$ is divided into $U^2_{2}$ and $U^{12}_{2}$. Then, the two private messages, $U^{1}_{1}$ and $U^{2}_{2}$, are encoded into private streams $s_{1}$ and $s_{2}$, respectively. From a common codebook, $U^{12}_{1}$ and $U^{12}_{2}$ are combined and encoded into one common stream $s_{12}$.
Without loss of generality, we assume that $s_{i}$ ($i \in \{1, 2, 12\}$) is randomly selected from a PAM constellation with zero mean and normalized range $[-1, 1]$. Let $\textbf{s}=\left[s_1,s_2,s_{12}\right]^T$ be the transmitted symbols vector, with $\mathbb{E}(\textbf{s}\textbf{s}^{T})=\textbf{\textit{I}}$. It is fruther  assumed that the non-linear response of the LED is compensated through digital pre-disposition (DPD) \cite{Stepniak2013}. To reduce MUI, a linear precoding matrix $\textbf{P}=\left[\textbf{p}_{1},\textbf{p}_{2},\textbf{p}_{12}\right]$ is considered, where $\textbf{p}_{i}=[p_{i,1} p_{i,2}]^T \in \mathbb{R}_{2 \times 1}$ is the precoding vector for the $i$th stream.  A DC bias $\textbf{d}_{DC} \in \mathbb{R}_{2 \times 1}$ is added in order to ensure positive signals,  which  is required by the LEDs. Hence, the transmitted signal, $\textbf{x} \in \mathbb{R}^+_{2 \times 1}$, can be written as   
\begin{equation}
\mathbf{x} = \left[x_1, x_2\right]^T= \mathbf{{P}} \mathbf{s}+\mathbf{d}_{DC}
  = \sum_{i \in \{1,2,12\}}\mathbf{p}_{i}s_{i}+\mathbf{d}_{DC},
\end{equation}
and the received signal at the $k$th PD, after optical to electrical conversion, is expressed by
\begin{equation}
y_{k}=\varsigma \zeta \mathbf{h}_{k}^{T} \mathbf{x}+{n}_{k}, \; \forall k \in \{1,2\}.
\end{equation} 
where $\varsigma$ is the conversion factor of any LED, $\zeta$ is the responsivity of any PD, $\mathbf{h}_{k}=[h_{k,1}, h_{k,2}]^T$ is the DC channel gain vector between the $k$th PD and the transmitting LEDs, where each element is expressed as given in (\ref{eq:fixture_app}), and $n_k \sim \mathcal{N}(0,\sigma^2_k)$ is the additive white Gaussian noise (AWGN), representing the thermal and shot noise, with zero-mean and variance $\sigma^2_k$. Due to the low mobility of indoor users, we assume that the channel gains are constant during the transmission, and that perfect CSI is available at the transmitter in order to accurately design the precoding matrix $\textbf{P}$. 
 

Recalling that the optical intensity must be a real value and nonnegative, in order to realize IM/DD,  the precoding matrix $\textbf{P}$ has to be carefully crafted. First, the input signals to the $l${th} transmitter LED must be positive \cite{Shen2016}, i.e., 
\begin{equation}
\label{eq:Tx}
x_{l}=\sum_{i \in \{1,2,12\}} p_{l,i} s_{i} + d_{DC} \geq 0, \; \forall l \in \{1,2\}.
\end{equation}
Taking into account the worst case, i.e., the left side of (\ref{eq:Tx}) is minimal, then $p_{l,i} \, s_i=-|p_{l,i}|$, $\forall l\in \{1,2\},i \in \{1, 2, 12\}$. Thus, the first constraint can be rewritten
\begin{equation}
\label{eq:c1}
\sum_{i\in\{1,2,12\}} |p_{l,i}|= L_1(\mathbf{p}_l) \leq d_{DC}, \; \forall l \in \{1,2\}.
\end{equation}
Moreover, the transmitted power of each LED is limited by a maximal value $P_{\rm{max}}$. This limitation is needed in order to alleviate the over-heating and dynamic range issues.
Thus, it can be integrated into a power constraint by maximizing the left side of the inequality in (\ref{eq:Tx}), i.e., $p_{l,k} \, s_k=|p_{l,k}|$. The following constraint is then deduced as 
\begin{equation}
\label{eq:c2}
\sum_{i\in \{1,2,12\}} |p_{l,i}|=L_1(\mathbf{p}_l) \leq P_{\rm{max}}-d_{DC},\; \forall l \in \{1,2\}.
\end{equation}
By combining both constrains (\ref{eq:c1})-(\ref{eq:c2}), the resulting precoding constraint can be expressed as
\begin{equation}
\label{eq:c3}
L_1(\mathbf{p}_l) \leq \varepsilon=\text{min}\left(d_{{DC}},P_{\rm{max}}-d_{{DC}}\right), \; \forall l \in \{1,2\}.
\end{equation}

Signal detection at the $k$th user can be performed using  the following MMSE equalizer \cite{Ma2013}
\begin{equation}
g_i = \mathbf{p}^T_i \mathbf{h}_k \left(1+\mathbf{p}^T_i \,\mathbf{h}_k \, \mathbf{h}_k^T\, \mathbf{p}_k \right)^{-1},\; \forall i
\end{equation}
yielding the following estimated stream
\begin{equation}
	      \hat{s}_i= g_i \; y_k,\; \forall i.          
\end{equation}
At user $k$, signals decoding is performed as follows. First, user $k$ decodes the common signal $s_{12}$ while treating the other signals as noise, i.e., apply MMSE equalizer, $g_{12}$. 
Hence, the received SINR at the $k$th user, for the common signal, can be given by \cite{Yin2016}
\begin{equation}
\label{eq:commom}
\gamma^{12}_k = \frac{\left(\mathbf{h}^T_k \mathbf{p}_{12}\right)^2}{\left(\mathbf{h}^T_k \mathbf{p}_1\right)^2 + \left(\mathbf{h}^T_k \mathbf{p}_2\right)^2+ \hat{\sigma}^2_k}, \; \forall k \in \{1,2\},       
\end{equation}
where $\hat{\sigma}^2_k=\sigma_k^2 / \left( \varsigma \zeta \right)^2$ is the normalized received noise power. For the sake of simplicity, we assume that $\varsigma \zeta=1$, thus $\hat{\sigma}_k^2=\sigma_k^2$. 
Then, the effect of common signal is removed using SIC. This allows for the detection of the private signal. Similarly, user $k$ attempts to decode its private message $s_k$ while treating  other user's private signal as noise. Consequently,    
the received SINR at user $k$, for its private signal, can be written as
\begin{equation}
\label{eq:private}
\gamma^{k}_k = \frac{\left(\mathbf{h}^T_k \mathbf{p}_{k}\right)^2}{\left(\mathbf{h}^T_k \mathbf{p}_{\bar{k}}\right)^2 +  {\sigma}^2_k}, \; \forall (k,\bar{k}) \in \{(1,2),(2,1)\}.
\end{equation}
The achieved data rate at user $k$ can be given as follows
\begin{equation}
R^{12}_k=\text{log}_{2}(1+\gamma^{12}_k),\; \forall k\in \{1,2\}
\end{equation}
and
\begin{equation}
R_k=\text{log}_{2}(1+\gamma^{k}_k),\; \forall k\in \{1,2\}
\end{equation}
where $R_k^{12}$ and $R_k^k$ are the data rates for the common and private signals, respectively.
In order to ensure successful decoding of the common stream $s_{12}$ at both users, the common rate shall not exceed $R_{12}=\text{min}(R^{12}_1,R^{12}_2)$. Rate boundaries for two-user rate splitting region can be obtained if $R_{12}$ is adequately shared between the two users, i.e., $R_{12}=\sum_{k=1}^2 R_{k,\rm{com}}$, where $R_{k,\rm{com}}$ is the $k$th user portion of the common rate. Consequently, 
the total achievable data rate of user $k$, denoted $R_{k,\rm{ov}}$, can be expressed by \cite{Mao2018}
\begin{equation}
\label{eq:c4}
R_{k,\rm{ov}}=R_{k,\rm{com}}+R_k, \; \forall k \in \{1,2\}.
\end{equation}
Although conventional precoders, such as ZF and ZF-DPC, are simple and can efficiently remove MUI, they suffer from performance degradation at low SNR values. Consequently, there is a need for optimal precoding in order to maximize an objective function, e.g., sum rate, weighted sum rate, proportional fairness, or max-min fairness \cite{Pham2017}, under per-LED transmit power constraints and QoS requirements. Inspired by  the MMSE precoding method presented in \cite{Nguyen2014},  we maximize the WSR of a MU-MISO VLC system. This method is proven to be superior to conventional precoding techniques. \\
For a given weights vector $\textbf{w} =[w_1,w_2 ]$, the WSR maximization problem (P1) can be expressed as follows 
\begin{subequations}
	\begin{align}
	\small
	\max_{\mathbf{P},\mathbf{R}_{\rm{com}}} & \quad 
	{R}(\textbf{w})={\sum_{k=1}^2}w_k \;R_{k,\rm{ov}} \tag{P1} \\
	\label{c1}
	\text{s.t.}\quad & L_1(\mathbf{p}_l) \leq \varepsilon, \; \forall l \in \{1,2\}  \nonumber \tag{P1.a}\\
	\label{c2} & \sum_{k=1}^2R_{k,\rm{com}} \leq R_{12},  \tag{P1.b}\\
	\label{c3} & \mathbf{R}_{\rm{com}}\geq \mathbf{0},  \tag{P1.d}
	\end{align}
\end{subequations}

\noindent
where $\textbf{R}_{\rm{com}}=[R_{1,\rm{com}},R_{2,\rm{com}}]$ is the common rate vector, and $P_t$ is the transmit power. (P1) is non-convex due to the presence of variables $\textbf{p}_k$ ($k \in \{1,2\}$) in the denominator of the SINR expressions (\ref{eq:commom})-(\ref{eq:private}). Thus, its solution is not straight-forward. Similar to \cite{Christensen2008}, we opt for problem reformulation, where the objective becomes the minimization of the weighted MMSE (WMMSE), and is achieved by jointly optimizing the WMMSE precoding vectors and MSE weights. 
To obtain a local optimum, we utilize  alternating optimization (AO) \cite{Christensen2008}, where in order to converge to a maximum WSR, it alternates between WMMSE precoding design and MSE weights design. For further details on the AO procedure, we refer the reader to Sections IV and V in \cite{Christensen2008}.\\
Finally, the reformulated problem can be solved using  optimization software such as CVX in MATLAB\cite{CVX}. 
It is to be noted that the AO algorithm converges faster and with better performances than other types of precoding optimization algorithms. 
However, its complexity increases if the number of users is larger than two. 
\subsection {NOMA and SDMA as Special Cases of RSMA} 
As we mentioned earlier, RSMA is a generalized MA scheme, where NOMA and SDMA are special cases. To implement SDMA from RSMA, the common stream is allocated null power, and each user's message is encoded into a private stream only. Hence, the transmitted signal in this case is
\begin{equation}
\mathbf{x}=\mathbf{P} \mathbf{s}+\mathbf{d}_{DC}=\sum_{i\in\{1,2\}}\mathbf{p}_i s_i+\mathbf{d}_{DC},
\end{equation}
and the received SINR at each user simplifies into (\ref{eq:private}).

\begin{table*}[t]
\scriptsize
\centering
\caption{SDMA vs. NOMA vs. RSMA}
\begin{tabular}[l] {|p{1.5cm}|p{2cm}|p{2cm}|p{1.5cm}|p{2cm}|p{1.5cm}|p{2.5cm}|p{2cm}|}
\hline
\textbf{MA technique}&\textbf{Transmitter design}&\textbf{Receiver design}&\textbf{Users' signals separation }&\textbf{Network load}&\textbf{Channel conditions}&\textbf{Transmitted Streams}&\textbf{Complexity}  \\ \hline \hline 					
SDMA & Linear precoding of a number of streams less or equal to the number of LEDs & Interference from other users' streams is treated as noise & At transmitter & Underloaded (number of users less or equal to number of LEDs) & 	Similar and semi-orthogonal channel gains &	Users’ messages are all encoded into private streams &
Complex scheduler and encoder. No SIC at receiver
 \\ \hline
NOMA & Linear precoding and superposition coding & SIC (number of SIC layers equal to number of served users-1) & At receiver &	Overloaded (number of users exceeds number of LEDs)&	Different
channel gains & Users' messages with strong channels encoded into private streams, while users' messages with weak channels are encoded into common streams.	
& Complex scheduler, encoder and receiver\\ \hline
RSMA & Linear precoding and rate splitting  (number of streams equal to number of served users+number of combined common messages between users) & SIC (number of SIC layers equal to number of served users in a spatially-separated group+number of combined common messages between users in this group) & At both transmitter and receiver &	Underloaded and Overloaded & Any channel gain &	Messages are split into common and private streams & Complex scheduler, encoder and receiver  \\ \hline
\end{tabular}
\label{Table2}
\end{table*}

Similarly, NOMA can be obtained from RSMA by encoding one of the users' messages as a private stream, i.e., the user with the strongest channel, and  the signal of the second user is encoded into a common stream. Assuming that user 1 has the strongest channel gain, then the transmitted signal in this case can be written as 
\begin{equation}
\mathbf{x}=\mathbf{P}\mathbf{s}+\mathbf{d}_{DC}
= \sum_{i \in \{1,12\}}\mathbf{p_i} s_i+\mathbf{d}_{DC},
\end{equation}
and the associated SINRs are given by   
\begin{equation}
\gamma_1^1 = \frac{(\mathbf{h}^T_1 \mathbf{p}_{1})^2}{{\sigma}^2_1},    
\end{equation}
 and
\begin{equation}
\gamma_2^{12} = \text{min}\bigg(  \frac{(\mathbf{h}^T_1 \mathbf{p}_{12})^2}{ (\mathbf{h}^T_1 \mathbf{p}_1)^2+ {\sigma}^2_1}\;,\;\frac{(\mathbf{h}^T_2 \mathbf{p}_{12})^2}{(\mathbf{h}^T_2 \mathbf{p}_1)^2+ {\sigma}^2_2}\bigg).  \end{equation}
It is worth mentioning that the flexibility of RSMA comes at the expense of a slightly higher encoding complexity at the transmitter. 
Table \ref{Table2} summarizes the three MA schemes and the main differences between them.

\begin{figure*}[t]
     \begin{minipage}{0.5\linewidth}
     \includegraphics[width=3in]{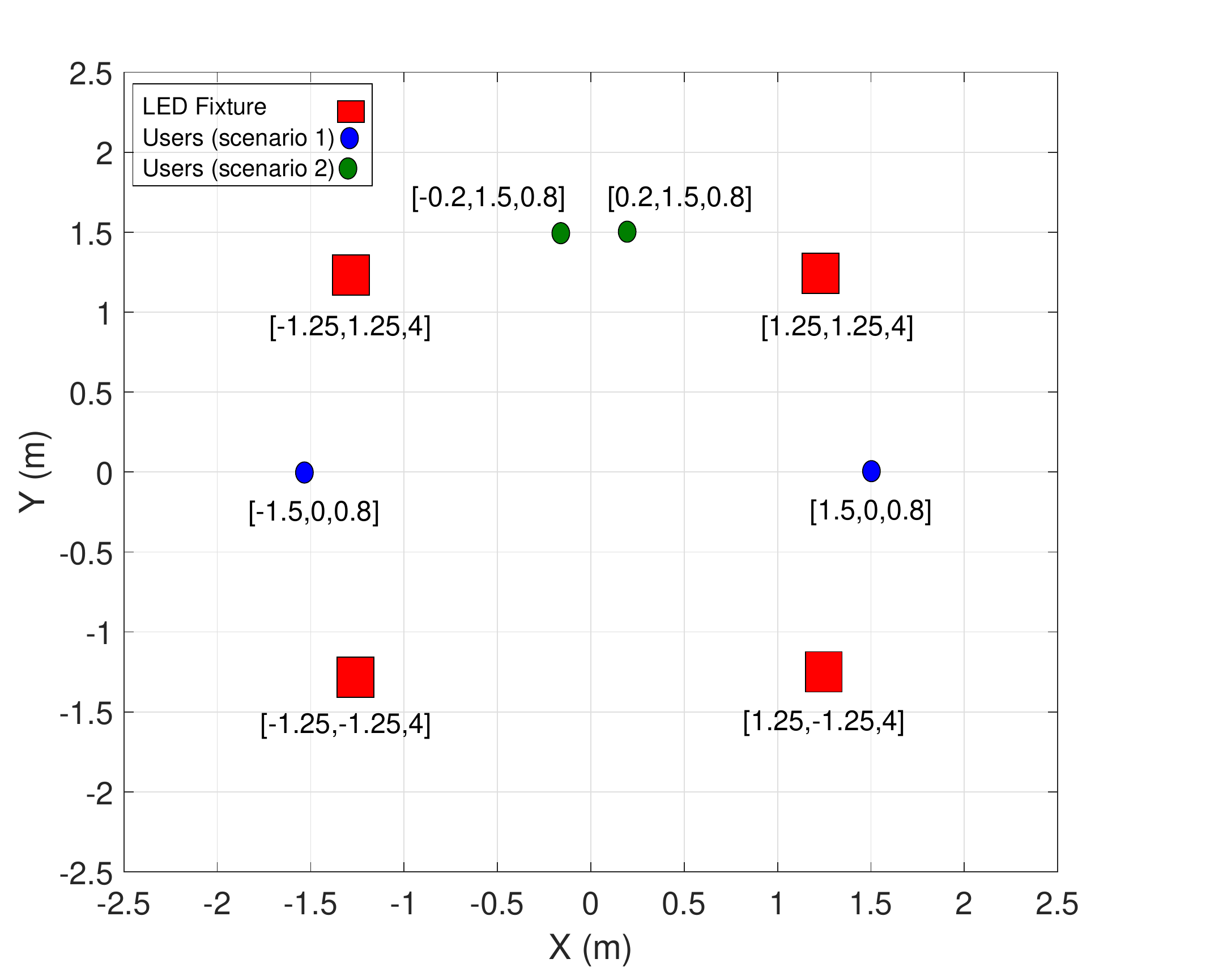}
       \caption{Room configuration and users' scenarios (4 LEDs).}
    \label{Fig:4LEDs}
     \end{minipage}
     \hfill
     \begin{minipage}{0.5\linewidth}
   \includegraphics[width=3in]{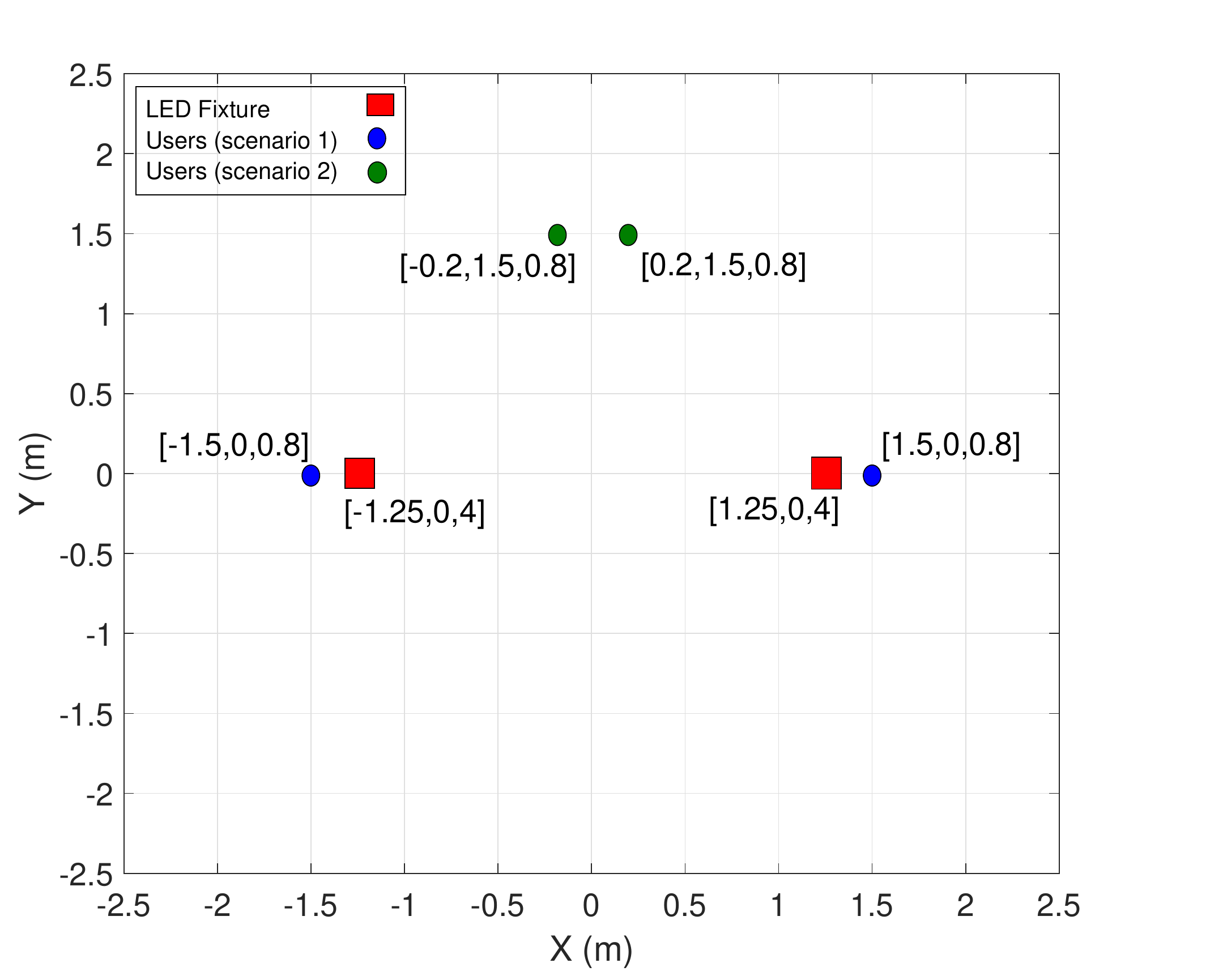}
	\caption{Room configuration and users' scenarios (2 LEDs).}
	\label{Fig:2LEDs}
     \end{minipage}
   \end{figure*}

\section {Performance Study}
We present in this section different scenarios for the application of RSMA in VLC systems, where we investigate their performance, in terms of WSR, and then compared to SDMA and NOMA. Moreover, we study the impact of changing  users' locations within an indoor space on WSR. 


\begin{figure}[t]
\centering
\includegraphics[width=3in]{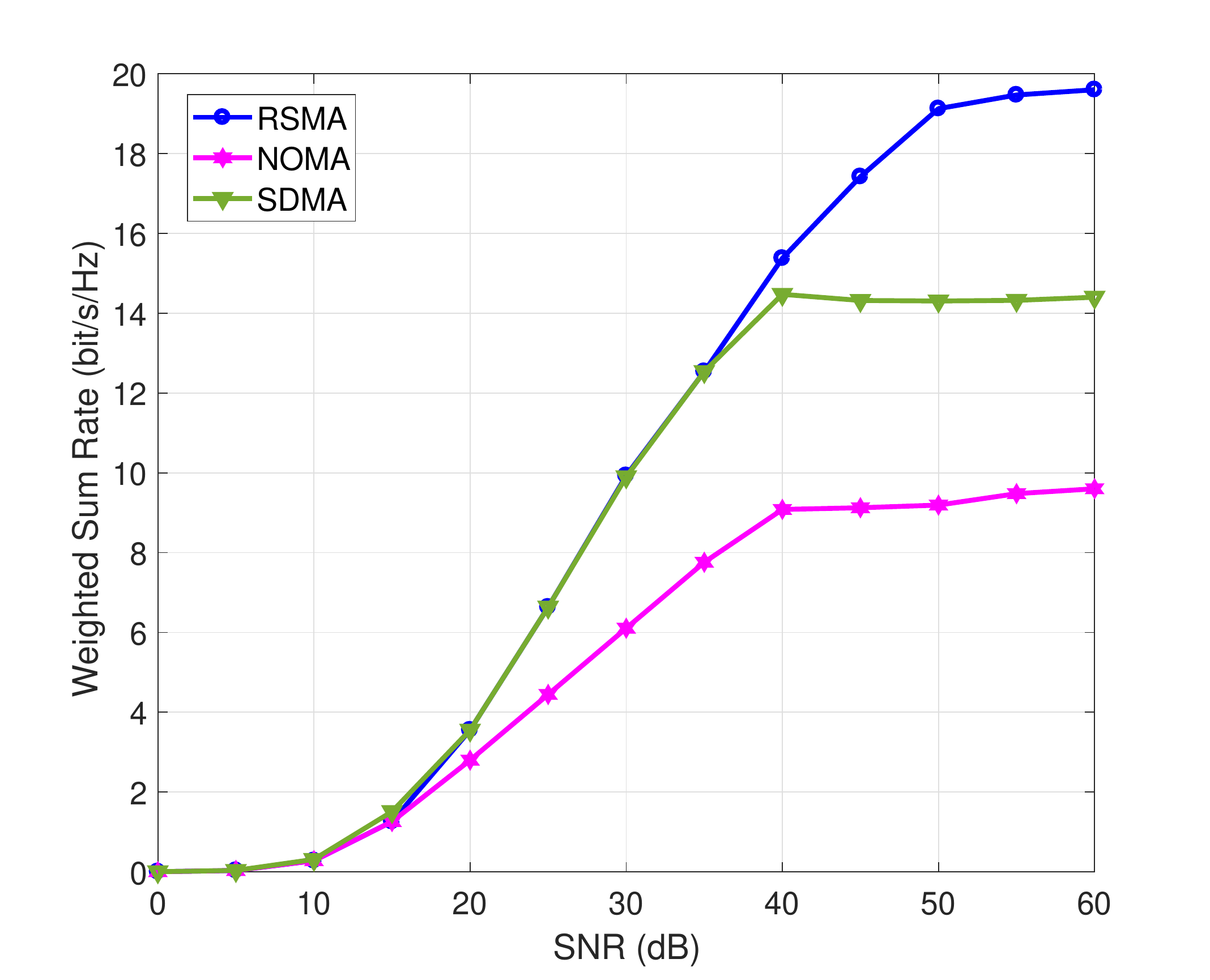}
\caption{WSR vs. SNR (Scenario 1, 4 LEDs).}
\label{Fig:Sim1}
\end{figure}

\begin{table}[t]
\scriptsize
\centering
 \caption{Simulation parameters.}
\begin{tabular} [c]{|p{3cm}|p{1.5cm}|p{2cm}|}
\hline 
\textbf{Parameter}&\textbf{Symbol}&\textbf{Value} \\ \hline \hline
Number of LEDs per fixture & $Q$ &	$3600\; (60\times60)$\\ \hline
LED beam angle& $\varphi_{1/2}$ &	$60^o$\\ \hline
PD area& $A_k$ ($k=1,2$) &	$1\; cm^2$\\ \hline
Refractive index of PD& $n$ & $1.5$\\ \hline
Gain of optical filter&	$T_s(\phi_{k,i})$ ($k=1,2$) &1\\ \hline
FoV of PD & $\phi_c$	& $60^o$\\ \hline
\end{tabular}
\label{TableIII}
\end{table}

We consider a RSMA-based MU-MISO VLC system, where two  single-PD users are served by two or four LED fixtures in a room of size $5\times5\times4 \;m^3$. 
The room configurations with the users' scenarios are detailed in Figs. \ref{Fig:4LEDs}--\ref{Fig:2LEDs} as follows. In both Figs., two users' location scenarios are considered. In the first (blue circles), users are located in the middle space of the room with a separation of 3 m,  whereas in the second (green circles), users are located in the top of the room, with a smaller separation of 0.4 m. Between the two Figs., the number and locations of LEDs is varied from 4 to 2. 
All coordinates are expressed in the 3D-space system. Also, we assume that the optical devices characteristics are as the ones in \cite{Ma2013}, while the two users are allocated equal priority, i.e., $w_1=w_2=\frac{1}{2}$ in the objective function of (P1). Since the noise power is assumed unitary, then SNR designates the transmit power $P_t$. The remaining parameters are detailed in Table \ref{TableIII}.


\begin{figure}[t]
\centering
\includegraphics[width=3in]{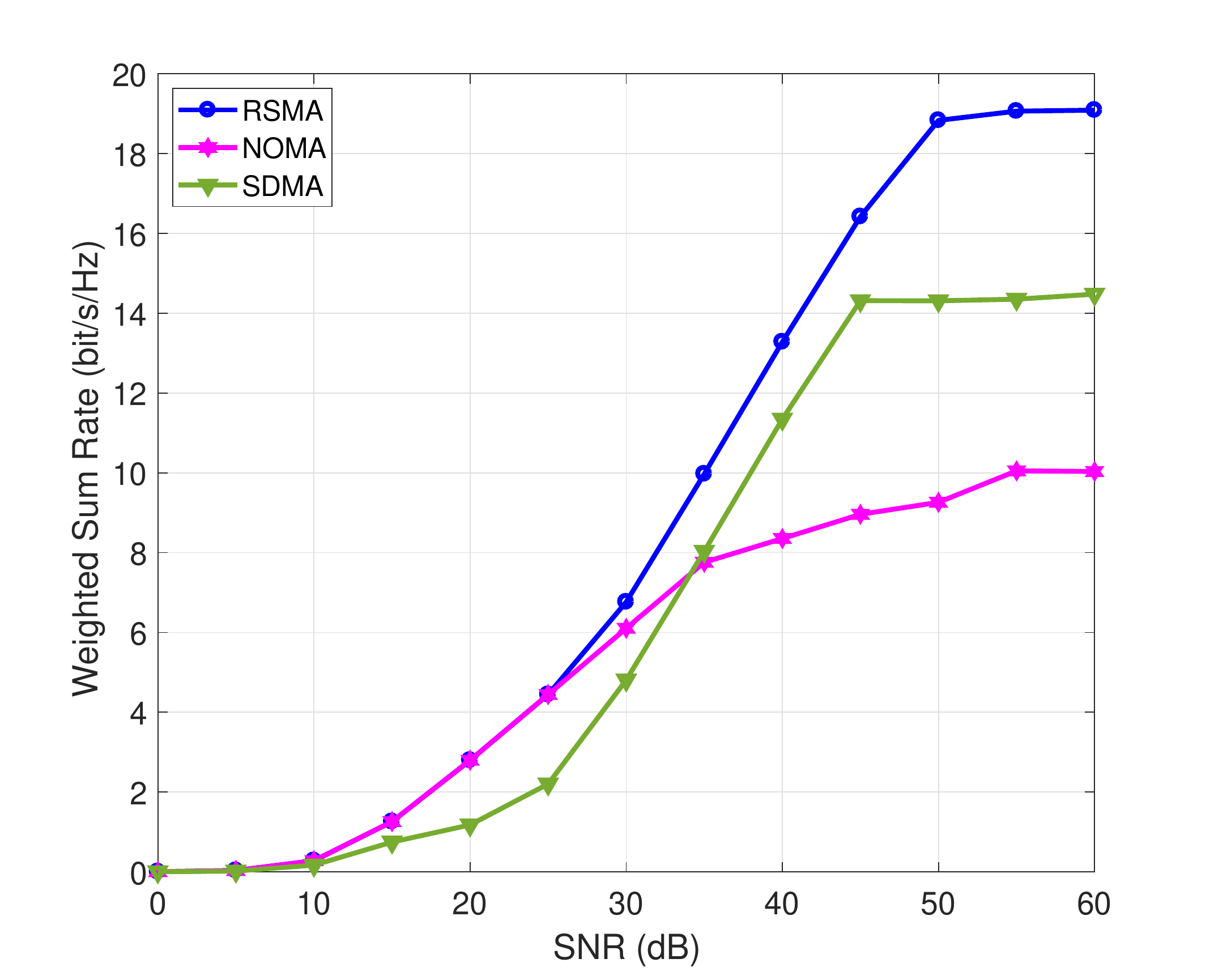}
\caption{WSR vs. SNR (Scenario 2, 4 LEDs).}
\label{Fig:Sim2}
\end{figure}

Fig. \ref{Fig:Sim1} shows the WSR performance for RSMA, NOMA and SDMA, for ``Scenario 1, 4 LEDs". It can be seen that RSMA outperforms both NOMA and SDMA, especially at high SNR. In addition, SDMA performs better than NOMA, since the number of transmitter LEDs is larger than the number of users, allowing efficient management of MUI. However, SDMA performs worse than RSMA due to the difficult channels alignment between users, caused by the nature of the VLC channel. In Fig. \ref{Fig:Sim2}, the same comparison is made for ``Scenario 2, 4 LEDs". With a smaller separation between users, channels are more correlated. This accentuated correlation is reflected in the performance. For instance, using RSMA, WSR=13 bits/s/Hz (RSMA) at SNR=40 dB, compared to WSR=15.5 bits/s/Hz in Fig. \ref{Fig:Sim1}.  
Nevertheless, the performance of RSMA still exceeds that of both NOMA and SDMA. 
At low SNR (below 35 dB), NOMA outperforms SDMA. Indeed, NOMA is able to distinguish the different users using precoding and SIC receivers. However, at higher SNRs, this procedure is less interesting, and direct beamforming using SDMA becomes more efficient. Consequently, NOMA is favored at low SNR for low users separation, however, SDMA is more performing at high SNR. 


\begin{figure}[t]
\centering
\includegraphics[width=3in]{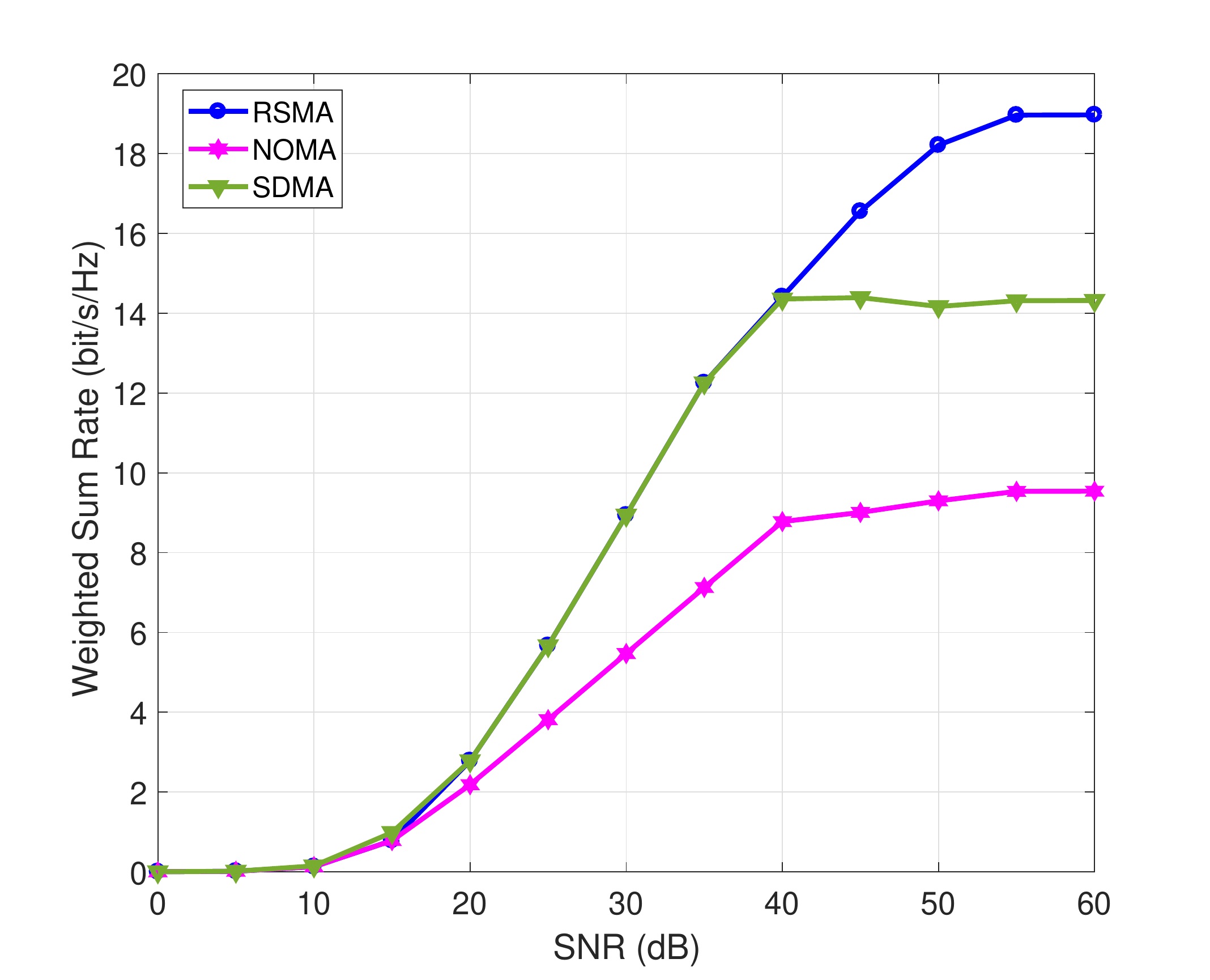}
\caption{WSR vs. SNR (Scenario 1, 2 LEDs).}
\label{Fig:Sim3}
\end{figure}

\begin{figure}[t]
\centering
\includegraphics[width=3in]{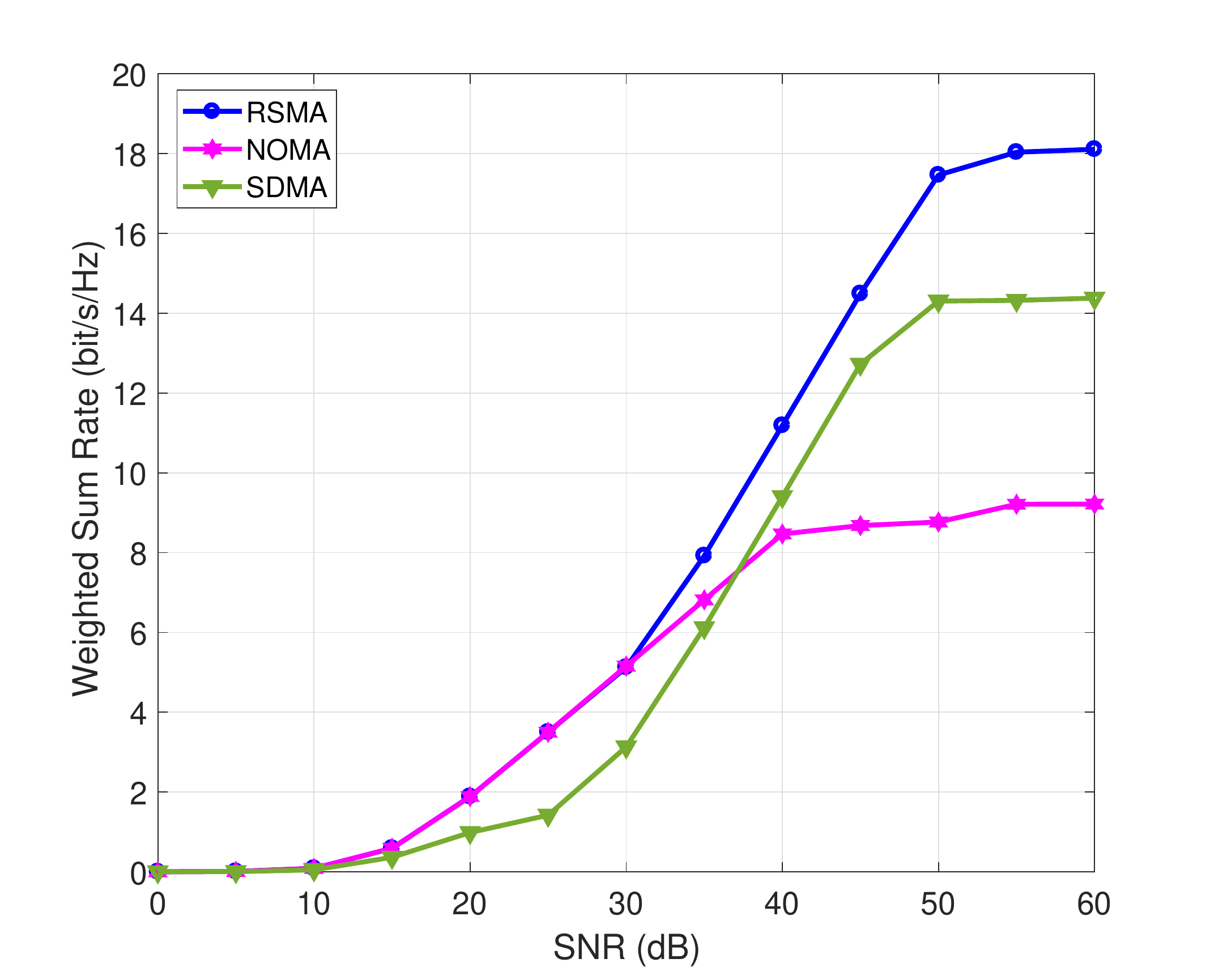}
\caption{WSR vs. SNR (Scenario 2, 2 LEDs).}
\label{Fig:Sim4}
\end{figure}

In Figs. \ref{Fig:Sim3}-\ref{Fig:Sim4}, we consider the same scenarios, but with 2 LEDs only. Similar to the previous results, the superiority of RSMA over the other techniques, in terms of WSR, is clearly illustrated. SDMA's performance is slightly degraded due to the smaller number of LEDs. Similarly to Fig. \ref{Fig:Sim2}, in Fig. \ref{Fig:Sim4} SDMA performance is degraded at low SNR (below 36 dB) compared to NOMA, but outperforms the latter as SNR increases.

\begin{figure}[t]
\centering
\includegraphics[width=3.25in]{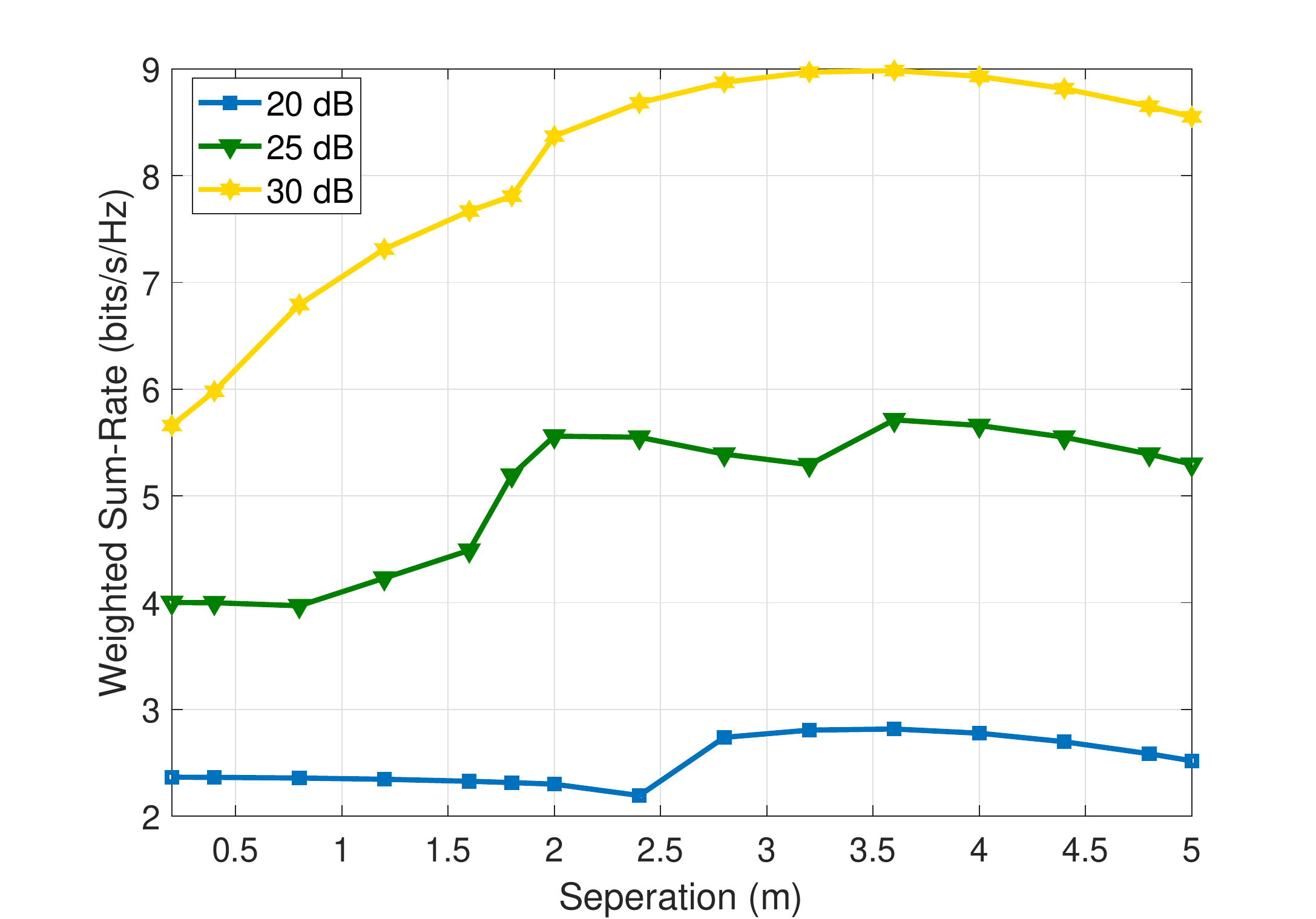}
\caption{Performance of RSMA for different users’ locations/separations and SNRs.}
\label{Fig:Sim5}
\end{figure}

Fig. \ref{Fig:Sim5} illustrates the users' locations impact on the WSR performance of the RSMA scheme. We considered the room setup of 2 LEDs, and two users initially located in the middle of the room. From there, the first and second users travel to the east and west walls at the same constant speed, respectively. Thus, their physical separation increases until reaching its maximum 5 m. It can be seen that WSR varies with the separation, until a maximum value is achieved for a separation equal to 3.6 m. This corresponds to users locations [-1.8,0,0.8] and [1.8,0,0.8], where correlation between channels is low, but users are very close to one of the serving LEDs to capture maximal power. However, as this separation increases above 3.6 m, WSR degrades due to longer distances between users and LEDs. It can be seen that these optimal users' locations are the same for different SNRs. 
Consequently, designing indoor spaces using RSMA-VLC requires a careful consideration of the LEDs' and users' locations.

\section{Open Issues and Research Directions}

In this paper, we presented the first comprehensive approach to the utilization of RSMA in VLC systems. It has been shown that RSMA is a powerful MA scheme that can provide high data rates and reliable communications in  VLC systems. However, several associated issues need to be addressed and analyzed for the practical realization of RSMA-VLC.
For instance, the impact of 
the non-linear distortion caused by the different circuits components, such as LEDs, PDs, and analog/digital and digital/analog converters has to be investigated. Moreover, as a novel MA technique, more efforts are required to study the design of the PHY and MAC layers. Hence, different performance metrics, modulation and coding schemes, and security issues, are open research problems in the RSMA-VLC. Additionally, optimal precoding and power allocation for RSMA-VLC are still open for investigation, where new linear or non-linear techniques can be proposed and optimized. Moreover, the current literature has mainly the Gaussian noise assumption, but neglected  the effect of ambient light, which can cause significant degradation in performance.

Other current assumptions include: the receiver is always pointing upward, a line-of-sight is always available and CSI is perfect. However, this may not be the case in  practical scenarios, where the receiver can be differently oriented, the VLC link may experience shadowing and/or blockage, and CSI is imperfect. Consequently, the design and performance evaluation of RSMA-VLC systems that take into account these practical concerns have to be studied.
Innovative solutions  to circumvent the absence of a line-of-sight include enabling optical cooperative communications or device-to-device (D2D).  Optical cooperation among VLC transmitters guarantees reliable transmission to users in a specific area \cite{Pham2019},  whereas, in D2D communications, users with strong VLC links may help to relay data to users with blocked VLC channels \cite{Raveendran2019}. In the design of such systems, taking into consideration the different users' QoS may lead to improved performances.


Finally, the analysis of massive MIMO RSMA-VLC systems  is another interesting open research problem.


\section {Conclusion}
In this paper, several  MA techniques were discussed in terms of  their advantages and limitations. Then, we presented the first comprehensive approach for the integration of RSMA technique with VLC. In particular, we proposed a RSMA- VLC framework, where the expressions of SINR and WSR are obtained for a two-user MISO VLC system.   
Channel knowledge at the transmitter was exploited in the precoder design in order to  mitigate MUI and maximize the WSR performance. Results showed the flexibility of RSMA and its generalization over NOMA and SDMA at  slightly increased design complexity. It has been proved through simulation results that RSMA-VLC outperforms both techniques in terms of WSR. It has been further shown that   RSMA is robust against channel correlation. Hence, it can be seen as a strong multiple access candidate for VLC networks. Finally, a number of open issues and research directions, linked to MIMO RSMA-VLC,  have been presented and discussed.



\bibliographystyle{IEEEtran}
\bibliography{tau}

\end{document}